\documentclass[12pt]{spieman}  
\pdfoutput=1
\usepackage{amsmath,amsfonts,amssymb}
\usepackage{graphicx}
\usepackage{setspace}
\usepackage{tocloft}
\usepackage{amsmath, siunitx}
\usepackage{caption}
\usepackage{subcaption}
\ProvideDocumentCommand\unit{om}{\si[#1]{#2}}
\ProvideDocumentCommand\qty{omm}{\SI[#1]{#2}{#3}}
\ProvideDocumentCommand\qtylist{omm}{\SIlist[#1]{#2}{#3}}
\ProvideDocumentCommand\qtyrange{ommm}{\SIrange[#1]{#2}{#3}{#4}}
\DeclareSIUnit\counts{counts}
\DeclareSIUnit\FoV{FoV}

\title{Characterization of a TES-based Anti-Coincidence Detector for Future Large Field-of-View X-ray Calorimetry Missions}

\author[a,b,*]{Samuel V. Hull}
\author[a,c]{Joseph S. Adams}
\author[a]{Simon R. Bandler}
\author[d]{Matthew Cherry}
\author[a]{James A. Chervenak}
\author[a]{Renata Cumbee}
\author[d]{Xavier Defay}
\author[e]{Enectali Figueroa-Feliciano}
\author[a,f]{Fred M. Finkbeiner}
\author[a,c]{Joshua Fuhrman}
\author[a]{Richard L. Kelley}
\author[d]{Christopher Kenney}
\author[a]{Caroline  A. Kilbourne}
\author[d]{Noah Kurinsky}
\author[a]{Jennette Mateo}
\author[a,g]{Haruka Muramatsu}
\author[a]{Frederick S. Porter}
\author[a,c]{Kazuhiro Sakai}
\author[h]{Aviv Simchony}
\author[a]{Stephen J. Smith}
\author[h]{Zoe Smith}
\author[a,c]{Nicholas A. Wakeham}
\author[a]{Edward J. Wassell}
\author[a,i]{Sang H. Yoon}
\author[j]{Betty A. Young}

\affil[a]{NASA Goddard Space Flight Center, 8800 Greenbelt Rd., Greenbelt, MD 20771}
\affil[b]{University of Maryland, Department of Astronomy, College Park, MD 20742}
\affil[c]{University of Maryland, Baltimore County, Center for Space Sciences and Technology, Baltimore, MD 20742}
\affil[d]{SLAC National Accelerator Laboratory, 2575 Sand Hill Rd., Menlo Park, CA 94025}
\affil[e]{Northwestern University, Department of Physics and Astronomy, 2145 Sheridan Rd., Evanston IL 60208}
\affil[f]{Hexagon US Federal, 14291 Park Meadow Dr., Chantilly, VA 20151}
\affil[g]{Catholic University of America, Department of Physics, 620 Michigan Ave., N.E. Washington, DC 20064}
\affil[h]{Stanford University, Department of Physics, 382 Via Pueblo Mall, Stanford, CA 94305}
\affil[i]{Science Systems and Applications, Inc., 10210 Greenbelt Rd., Suite 600, Lanham, MD 20706}
\affil[j]{Santa Clara University, Department of Physics, 500 El Camino Real, Santa Clara, CA 95053}

\cftpagenumbersoff{figure}
\cftpagenumbersoff{table} 
\begin{document}

\maketitle

\begin{abstract}
Microcalorimeter instruments aboard future X-ray observatories will require an anti-coincidence (anti-co) detector to veto charged particle events and reduce the non-X-ray background. We have developed a large-format, TES-based prototype anti-coincidence detector that is particularly suitable for use with spatially-extended ($\sim$~\qty{10}{cm$^2$}) TES microcalorimeter arrays, as would be used for a future large field-of-view X-ray missions. This prototype was developed in the context of the Line Emission Mapper (LEM) probe concept, which required a $\sim$~\qty{14}{\square{\centi\meter}} anti-co detector with $>$ 95\% live time and a low-energy threshold below \qty{20}{\kilo\electronvolt}. Our anti-co design employs parallel networks of quasiparticle-trap-assisted electrothermal feedback TESs (QETs) to detect the athermal phonon signal produced in the detector substrate by incident charged particles. We developed multiple prototype anti-co designs featuring 12 channels and up to 6300 QETs. Here we focus on a design utilizing tungsten TESs and present characterization results. Broad energy range measurements have been performed (\qty{4.1}{\kilo\electronvolt} – \qty{5.5}{\mega\electronvolt}). Based on noise and responsivity measurements, the implied low-energy threshold is $<$~\qty{1}{\kilo\electronvolt} and a live time fraction of $>$~96\% can be achieved up to 5.5 MeV. We also find evidence of mm-scale-or-better spatial resolution and discuss the potential utility of this for future missions. Finally, we discuss the early development of a soild-state physics model of the anti-co towards understanding phonon propagation and quasiparticle production in the detector.

\end{abstract}


{\noindent \footnotesize\textbf{*}Samuel V. Hull,  \linkable{samuel.v.hull@nasa.gov} }

\section{Introduction}
\label{sect:intro} 

In the future, space observatories that fly very large X-ray microcalorimeter arrays will enable high-resolution spectroscopy and imaging over a wide field of view. This capability will open the door for new astrophysical studies; for example, by providing unprecedented insights into the physics of galaxy formation and evolution via mapping line emission from hot gas in the circumgalactic and intra-cluster media. Crucially, however, such advances rely on low non-X-ray background levels that can only be provided by a nearby anti-coincidence (anti-co) detector. 

\begin{figure}
\begin{center}
\begin{tabular}{c}
\includegraphics[width=\textwidth]{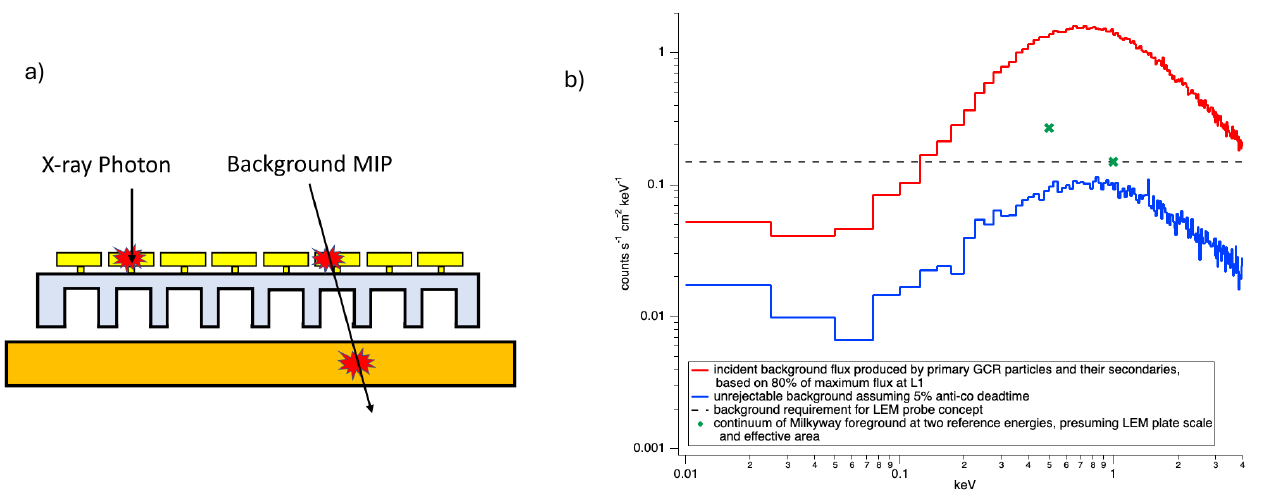}
\end{tabular}
\end{center}
\caption{(a) Cartoon of the anti-co concept, showing the anti-co (orange slab) positioned behind main microcalorimeter array (on back-etched substrate) to identify background MIP events. (b) Results of a Geant4 simulation for the LEM background (adapted from Athena/X-IFU simulations). The red curve is the total background (primaries $+$ secondaries) without any coincidence rejection. The blue curve shows the remaining unrejected background, assuming an anti-co with a \qty{20}{\kilo\electronvolt} low-energy threshold and a 5\% dead time. The green points represent the continuum component of the X-ray sky background at \qty{0.5}{\kilo\electronvolt} and \qty{1}{\kilo\electronvolt}, while the dashed horizontal line shows the LEM background requirement.}
\label{fig:background} 
\end{figure}

The non-X-ray background is composed of galactic cosmic-rays (GCRs) and secondary particles (primarily energetic electrons) produced via interactions of GCRs with the satellite and/or instrument housing. The GCR population is dominated by minimum ionizing particles (MIPs) (protons; $\sim$~90\%) that deposit energy in tracks as they pass through the detector, with most of the remainder being alpha particles ($\sim$~10\%)\cite{Lotti2021}. Energy depositions from GCRs and secondaries in the detector substrate are a potential noise source, but do not contribute to instrumental background\cite{Peille2020}; however, depositions in the array absorbers produce signals that are indistinguishable from valid X-ray events unless they can be vetoed via energy or coincidence. Unfortunately, the peak of the GCR energy deposition spectrum tends to lie squarely in the X-ray bandpass for which a given microcalorimeter instrument has been optimized. An anti-co detector is therefore typically placed directly behind the main array to flag coincident events that pass through both detectors so these events can be excluded from science analysis (see Figure \ref{fig:background} a). 

The required performance of an anti-co detector depends on the mission objectives and the design of an observatory, as the choice of microcalorimeter absorber, effective area of the telescope, and focal length determine the relative scaling of cosmic X-ray event rates (both signal and background and/or foreground terms) and the particle background. We focus on the concrete case of the Line Emission Mapper (LEM) probe concept, which would have flown a \qty{12.6}{\square\centi\meter} array of transition-edge sensors (TESs) covering a \qty{30}{\arcminute}~$\times$~\qty{30}{\arcminute} field of view (FoV) and operating in the \qtyrange{0.2}{2}{\kilo\electronvolt} energy bandpass\cite{Kraft2024, Bandler2023}. To achieve LEM’s science goals, a requirement was set for the non-X-ray background to be less than the sky background, corresponding to $<$ \qty{2}{\counts\per{\second\per{\kilo\electronvolt\per{\FoV}}}}\cite{Smith2023}. LEM’s thin absorbers optimized for the soft X-ray band would have seen a peak GCR energy deposition of $\sim$~\qty{0.8}{\kilo\electronvolt} at a rate $>$~\qty{14}{\counts\per{\second\per{\kilo\electronvolt\per{\FoV}}}} over the instrument bandpass\cite{Lotte_priv}. As illustrated in Figure \ref{fig:background} b), simulations show that a \qty{0.5}{\milli\meter} thick Si anti-co detector with a \qty{20}{\kilo\electronvolt} low-energy detection threshold and dead time $<$~5\% can reduce the background level below the LEM requirement. This plot shows the background energy deposition spectrum (combining primary GCRs and secondaries) in the LEM absorbers --- both with and without the anti-co, while also comparing to the X-ray sky background level. The residual background with the anti-co is composed of some unrejectable secondaries plus 5\% of the incident flux to account for the dead time when the anti-co is offline. With areal coverage extending slightly beyond the main array ($\sim$~\qty{14}{\square{\centi\meter}}) to account for events near the edge coming in at shallow angles, this anti-co would see a total count rate of roughly \qty{50}{\counts\per\second} with peak energy deposition of $\sim$~\qty{150}{\kilo\electronvolt}. While this discussion considered LEM in particular, similar anti-co configurations would be well-suited to other future large FoV mission concepts.

Microcalorimeter anti-co detectors using pin-diode technology have previously flown on Astro-H/SXS and XRISM/Resolve and could be scaled-up to the size required for a LEM-like mission\cite{Kilbourne2018}. However, scaling up would likely require segmentation to prevent capacitance from increasing, resulting in dead area between segments. In addition, as the anti-co will be placed physically close to the main detector array at the T0 temperature stage, the thermal, electrical, and mechanical designs of the instrument are significantly simplified by matching the anti-co detector and its readout to the rest of the instrument architecture. A TES-based anti-co design with superconducting quantum interference device (SQUID) readout is the best choice to match future TES microcaloriemter missions, and this aligns with the approach being taken for the Athena X-IFU anti-co\cite{Dandrea2022, Dandrea2024}. 

The X-IFU anti-co relies primarily on a thermal response to measure charged particle signals. While effective at X-IFU scales, scaling this approach to LEM-like sizes would be very challenging. By contrast, our anti-co design relies on an athermal technique that is common in the direct dark matter detection community to create large-area detectors with low energy threshold and high dynamic range\cite{Alkhatib2021}. The approach tiles quasiparticle-trap-assisted electrothermal feedback transition-edge sensors (QETs)\cite{Nam1996} over a large-area absorber with a long mean free path for phonons (e.g. high purity Si), sensing the athermal signal generated from energy depositions in the crystal substrate. These devices have demonstrated eV-scale thresholds and \qty{100}{\kilo\electronvolt} dynamic range in $\sim$~\unit{\square{\centi\meter}} detectors\cite{Ren2021}. The LEM anti-co was baselined with 12 channels, each consisting of a parallel network with $> 100$ QET unit cells.  We are investigating two versions of this anti-co design: one utilizing Mo/Au bilayer TESs, and one with tungsten TESs (W-TES).

This work concerns the characterization of a full-scale prototype anti-co detector that was originally developed in the context of LEM. In particular, our focus is on understanding the nature of the anti-co response to energy depositions, as well as on how the prototype performs on relevant anti-co performance metrics. To ground what follows, discussion of results is often benchmarked in comparison to LEM requirements; however, these results are widely applicable to other potential large-area, TES-based anti-co detectors. In Section \ref{sect:methods} we describe the design and experimental setup for testing of our prototype. Section \ref{sect:results} continues with presentation of the full characterization results, including key performance metrics like low-energy threshold and dead time, while also taking an early look at the potential spatial resolution that can be achieved with this design. In order to provide a more complete theoretical understanding of the anti-co, enable evaluation of design changes, and assist future calibration efforts, we have also developed a model of the anti-co with G4CMP (a Geant4 add-on package for phonon transport modeling)\cite{Kelsey2023}. Section \ref{sect:modeling} introduces the current status of the model, still under development, and presents preliminary results that can be compared to experimental data. Finally, we conclude in Section \ref{sect:conclusion} with a discussion of the relevance of these results for potential future missions and elaborate on next steps for continued development. 

\section{Methods}
\label{sect:methods}

\subsection{Detector Design}

Our baseline design builds on prior efforts developing small-format, TES-based anti-co prototypes. A single-channel, \qty{1}{\square{\centi\meter}} device with 121 TES unit cells previously demonstrated a threshold of $\sim$ \qty{200}{\electronvolt}\cite{Bailey2012, Busch2016}. We have now scaled up this design to a 12-channel, \qty{14}{\square{\centi\meter}} detector --- slightly larger than the main LEM array, to detect MIPs that pass through with shallow angles of incidence.  

\begin{figure}
\begin{center}
\begin{tabular}{c}
\includegraphics[width=.5\textwidth]{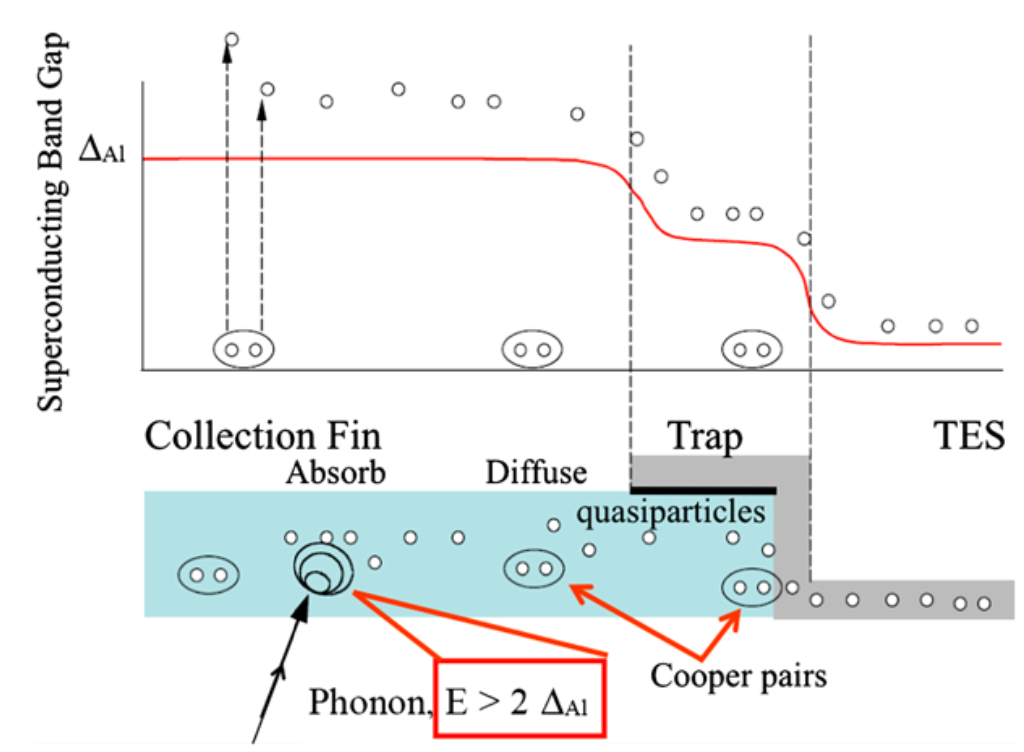}
\end{tabular}
\end{center}
\caption{Schematic of QET phonon collection/quasiparticle trapping process, showing a single TES unit cell with one Al fin. To improve yield, QETs are often fabricated with the Al collecting fins patterned after the TESs are defined, i.e. in an inverted stack compared to the geometry shown here.}
\label{fig:diagram} 
\end{figure}

This design relies on athermal phonon detection using quasiparticle trapping in QETs\cite{Irwin1995, Nam1996} fabricated on high-purity Si substrates. Figure \ref{fig:diagram} shows a diagram of the working principle. Each QET is composed of a long, thin TES connected to a number of Al collecting fins (see Table \ref{tab:params} for dimensions). Particle interactions in the bulk detector volume (Si) create athermal phonons that propagate to the wafer surface and get absorbed in the Al collecting fins. Phonons with energy greater than twice the Al superconducting energy gap break superconducting Cooper pairs in the fins, creating quasiparticles which then diffuse through the fins towards the TES. The quasiparticles are funneled to the band gap minimum at the voltage-biased TES before having their thermal energy measured as a change in TES current via a SQUID array. 

\begin{figure}
\begin{center}
\begin{tabular}{c}
\includegraphics[width=\textwidth, trim={0 2cm 0 0},clip]{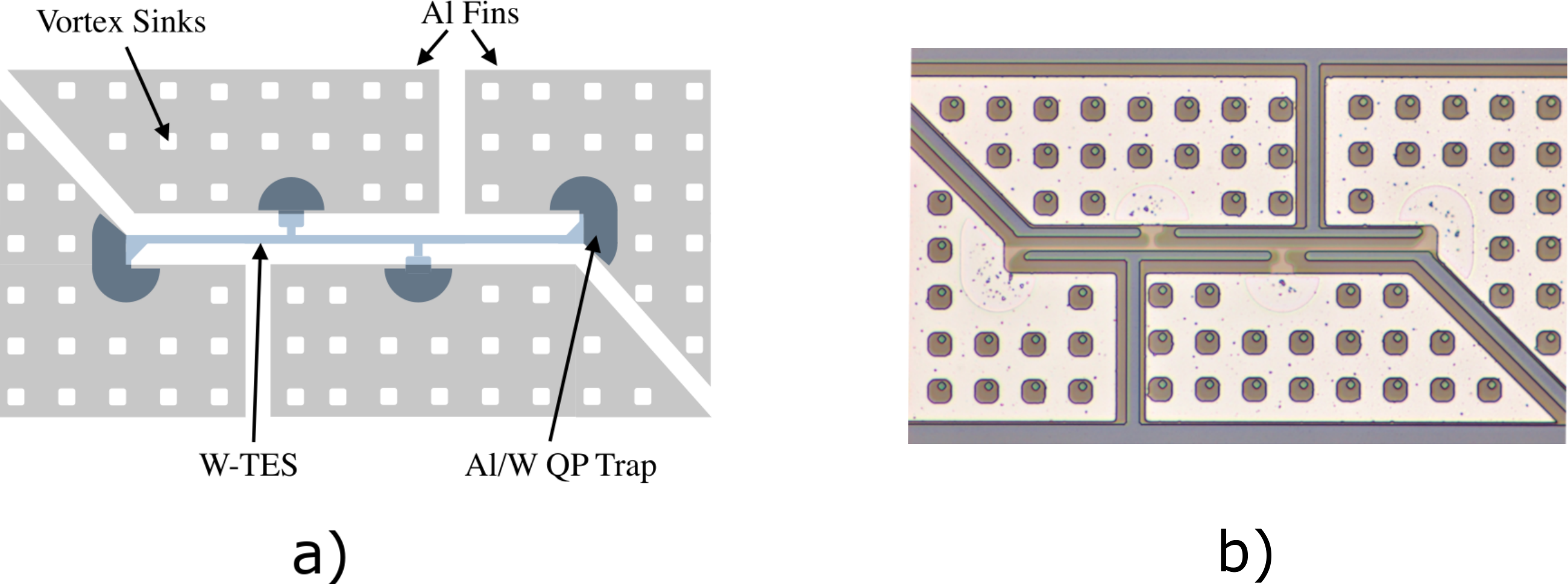}
\end{tabular}
\end{center}
\caption{(Left) Schematic of the W-TES QET unit cell geometry. The W-TES in light blue is the long, thin central object, which overlaps with the Al collection fins to form quansiparticle (QP) traps. The cutouts in the Al fins are vortex sinks designed to mitigate trapped flux. (Right) Photo of a QET unit cell on the prototype anti-co detector. The dark grey seen around the QET is amorphous silicon used as an interface in the fabrication process.}
\label{fig:unitcell} 
\end{figure}

We are currently developing two versions of this design, one utilizing Mo/Au bilayer TES unit cells (fabricated at NASA/GSFC) and one utilizing W-TES unit cells (fabricated at Stanford/SLAC); both use Al fins but with slightly different geometry. As designed, the W-TESs offer higher resistance than the Mo/Au TESs (hundreds of mOhms vs. $\sim$~10 mOhms). The two different TES designs also provide access to different parameter space, e.g. transition-temperature and thermal conductance. Full-scale prototypes of both designs have been fabricated and tested, but the work presented here focuses solely on the W-TES prototype. The design of the unit cell for this version is based on one used for ``HVeV'' (High-Voltage electron-volt) detectors for the SuperCDMS dark matter detection program \cite{Ren2021, Romani2018}. Figure \ref{fig:unitcell} shows the W-TES QET unit cell, while Table \ref{tab:params} summarizes the full set of QET design parameters used for this work. 

\begin{table}[ht]
\caption{Design parameters for the full-scale anti-co prototype.} 
\label{tab:params}
\begin{center}       
\begin{tabular}{l l l} 
\hline
\rule[-1ex]{0pt}{3.5ex}  Parameter & Description & Value \\
\hline\hline
\rule[-1ex]{0pt}{3.5ex}  $A_{\text{det}}$ & Active detector area & \qty{14}{\square\centi\meter} \\
\rule[-1ex]{0pt}{3.5ex}  $h_{\text{det}}$ & Detector thickness & \qty{0.5}{\milli\meter} \\
\rule[-1ex]{0pt}{3.5ex}  $N_{\text{ch}}$ & Number of channels & 12  \\
\rule[-1ex]{0pt}{3.5ex}  $N_{\text{QET}}$ & QETs per channel & $\sim$~530 \\
\rule[-1ex]{0pt}{3.5ex}  $l_{\text{TES}}$ & TES length & \qty{160}{\micro\meter}   \\
\rule[-1ex]{0pt}{3.5ex}  $w_{\text{TES}}$ & TES width & \qty{2.4}{\micro\meter}  \\
\rule[-1ex]{0pt}{3.5ex}  $V_{\text{TES}}$ & TES volume (per channel) & \qty{7390}{\cubic\micro\meter}  \\
\rule[-1ex]{0pt}{3.5ex}  $l_{\text{fin}}$ & Al fin length & \qty{60}{\micro\meter}  \\
\rule[-1ex]{0pt}{3.5ex}  $T_{\text{c}}$ & TES critical temperature & $\sim$~\qty{65}{\milli\kelvin}  \\
\rule[-1ex]{0pt}{3.5ex}  $R_{\text{n}}$ & Normal resistance (per channel) & $\qty{430}{\milli\ohm}$  \\
\hline 
\end{tabular}
\end{center}
\end{table} 

The detection thresholds required for LEM-like anit-co detectors are much larger than past HVeV performance, and thus some previous design constraints could be relaxed. For our full-scale W-TES prototype, we took a $\sim$~\qty{1}{\square\centi\meter}, two-channel HVeV design that acheives \qty{2}{\electronvolt} baseline resolution and \qty{120}{\kilo\electronvolt} dynamic range and expanded it to a twelve-channel design covering \qty{14}{\square\centi\meter}. We maintained a fixed number of QETs per channel ($N_{\text{QET}}$) and target $T_{\text{c}}$, with the sensor pattern simply spread out over a larger area and therefore scaling the channel area from \qty{0.5}{\square\centi\meter} to \qty{1.16}{\square\centi\meter} per channel. This means that, for a given channel, the intrinsic noise equivalent power is fixed, with the energy resolution only varying due to the increased channel count and slower signal collection time. In particular, the energy resolution (and to first order, also detection threshold) scale as:
\begin{equation}
    \sigma_E \propto T_{\text{c}}^3\sqrt{V_{\text{TES}} \cdot \tau_{\text{ph}}},
\end{equation}
whereas dynamic range (impacting the dead time) scales as:
\begin{equation}
    DR \propto \frac{1}{T_{\text{c}}^2}\sqrt{\frac{V_{\text{TES}}}{\tau_{\text{ph}}}},
\end{equation}
where $T_{\text{c}}$ is the TES critical temperature, $V_{\text{TES}}$ is overall volume of the TESs in all active channels, and $\tau_{\text{ph}}$ is the decay time of the measured phonon signal, which in an athermal device is determined by how long it takes for ballistic phonons to randomly hit the QETs and be absorbed in their aluminum collection fins\cite{Ren2021}. We expect $\tau_{\text{ph}}$ to increase directly proportional to the increase in channel area (a factor of $\sim 2.3$), while the channel count increasing from 2 to 12 with fixed $N_{\text{QET}}$ means $V_{\text{TES}}$ increases by a factor of 6. The larger channel count and slower $\tau_{\text{ph}}$ combine to give an expected overall degradation in energy resolution of under 4 (i.e., $\sqrt{6\cdot 2.3}$), such that we expect the small-signal resolution for the full design to be around \qty{10}{\electronvolt} and the low-energy threshold to be $<$~\qty{100}{\electronvolt}, assuming no additional signal efficiency loss. In a real device, other factors such as TES response time and large-signal nonlinearity may also be relevant. The expected dynamic range increases by roughly a factor of 2.6, such that we expect $>$~\qty{300}{\kilo\electronvolt} dynamic range and which should allow low deadtime to be achieved for typical GCR events.

\begin{figure}
\begin{center}
\begin{tabular}{c}
\includegraphics[width=\textwidth]{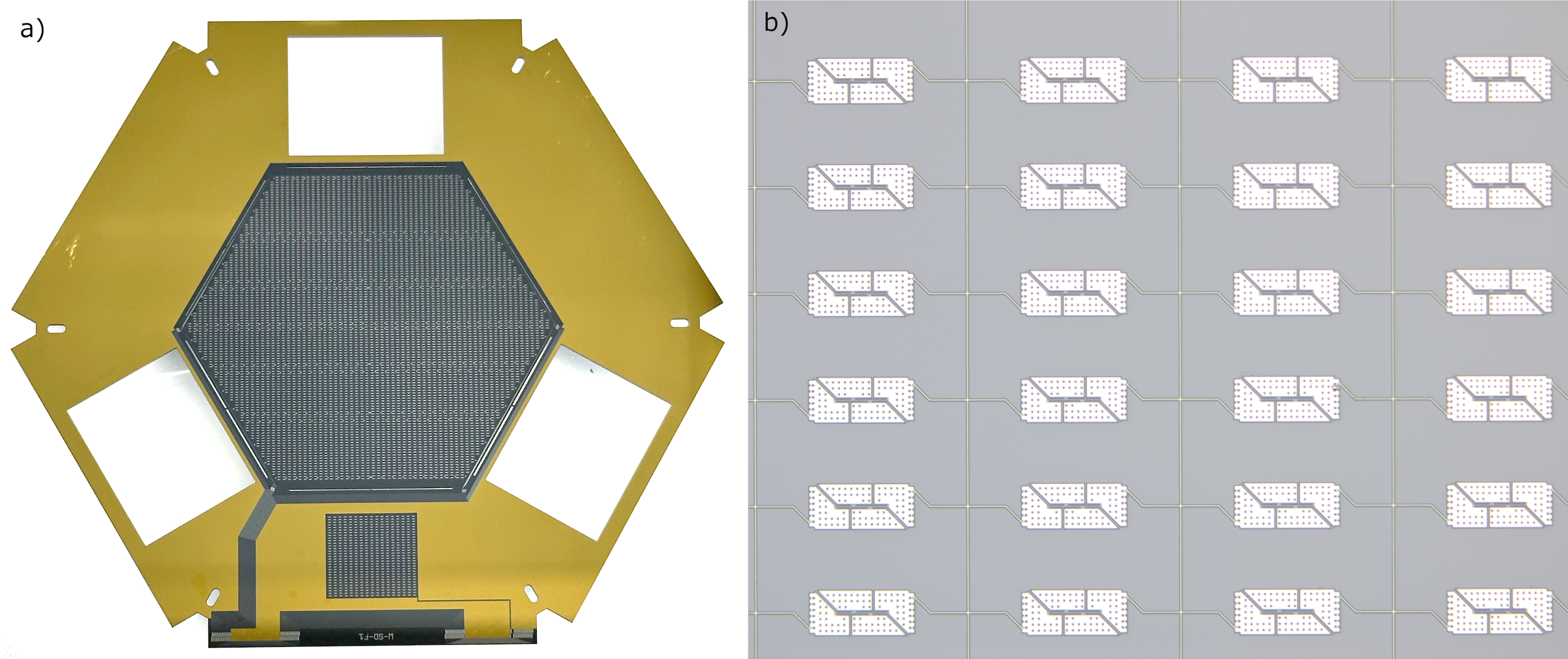}
\end{tabular}
\end{center}
\caption{Photos of the W-TES anti-co prototype. (a) The full wafer, showing all 12 channels partially surrounded by back-etched trenches. Channels are not visibly separated in the photo --- see Figure \ref{fig:masks} for channel layout. (b) One 6x4 block of QET cells, wired together in parallel and forming $\sim$~5\% of the area of a single channel. The Al bias wires of the parallel network are visible between cells.}
\label{fig:WTES_photo} 
\end{figure}

The full-scale W-TES anti-co prototype detector is shown in Figure \ref{fig:WTES_photo}. The detector contains $\sim$~6300 QET unit cells, divided into 12 roughly equal channels. The channels cover the central area of one side of a $\sim$ 0.5 mm-thick hexagonal shaped $\langle$100$\rangle$ Si wafer with flat-to-flat 
distances of \qty{79}{\milli\meter}. Figure \ref{fig:masks} shows the channel layout visualized over the detector area. All of the QETs in a channel are wired together in parallel with superconducting Al traces, forming a network with large areal coverage. The channel layout was chosen as a minimum viable solution containing 12 channels and keeping the QET pitch and number per channel consistent with previous HVeV designs. Layout optimization with regards to spatial resolution was not considered, but this may be a future area of investigation. Our prototype design features \qty{4}{\micro\meter}-wide Al traces in the interior of the detector. Unfortunately, a poor Al wet etch during fabrication resulted in degraded trace integrity in some locations, leading to some disconnected QETs in the array.

In order to reduce phonon losses, the central active area of the anti-co is partially isolated from the perimeter of the wafer using trenches etched into backside of the Si substrate. A new feature, to be implemented in the next W-TES fabrication run, will be to include a set of heaters just inside the trenches to enable independent temperature control of the anti-co while it is mounted to the same temperature stage as the main microcalorimeter array. This feature has been successfully implemented on prototype Mo/Au TES anti-co detectors in the past.

\subsection{Experimental Setup}

The anti-co prototype was measured in our ``960-pixel'' time-division multiplexing (TDM) platform that was originally designed to accommodate large-format ATHENA X-IFU wafer testing \cite{Sakai2023}. While capable of 24-column $\times$ 40-row TDM readout, here we made use of only 3 readout columns and a handful of rows: two columns of 3 channels each and one column of 6 channels. Each column contains a TDM-multiplexing chip with the first-stage SQUID amplifiers and TES-channel shunt resistors ($R_{\text{sh}}$; $\sim \qty{8}{\milli\ohm}$). The residual TES-circuit inductance was $< \qty{100}{\nano\henry}$, with no additional circuit inductance introduced. Channels within a single readout column are common biased, which means TES current cannont be individually optimized for each channel. The detector wafer is mounted to the top side of a hexagonal detector assembly, and a Kapton corner-turning flex with Nb superconducting traces carries the signals to the side-panel-mounted TDM and shunt chips. This full detector assembly was operated at a bath temperature ($T_{\text{b}})$ of \qty{53}{\milli\kelvin} in a He$^3$-backed adiabatic demagnetization refrigerator (ADR). Low-noise analog amplifiers provided by NIST and in-house built, commercial-off-the-the-shelf TDM digital electronics complete the readout chain\cite{Reintsema2009, Sakai2022}. The majority of measurements were performed with TDM using no more than 9 rows. Data was collected by saving simultaneous fixed-length pulse records for all channels for any triggered event.

We used precision masks and source collimators with a variety of sources to carefully study the detector response as a function of both event energy and location (see Section \ref{sect:pulses} for more detail). At the lower-energy end, florescence targets were used to generate characteristic X-rays ranging from \qty{4.1}{\kilo\electronvolt} to \qty{32.2}{\kilo\electronvolt}. We also used a sealed Am-241 source to generate \qty{60}{\kilo\electronvolt} X-rays. Finally, we used an unsealed Am-241 alpha-particle source mounted directly behind the wafer to deposit \qty{5.5}{\mega\electronvolt} and characterize detector response near the very top end of the expected energy range of the anti-co. It should also be noted that the large area of the detector resulted in a relatively high laboratory background rate of $\sim$~\qty{0.5}{\counts\per\s} for cosmic muons and terrestrial background gamma-rays, with typical deposition energies of \qtyrange{100}{200}{\kilo\electronvolt}.

\begin{figure}
\begin{center}
\begin{tabular}{c}
\includegraphics[width=\textwidth]{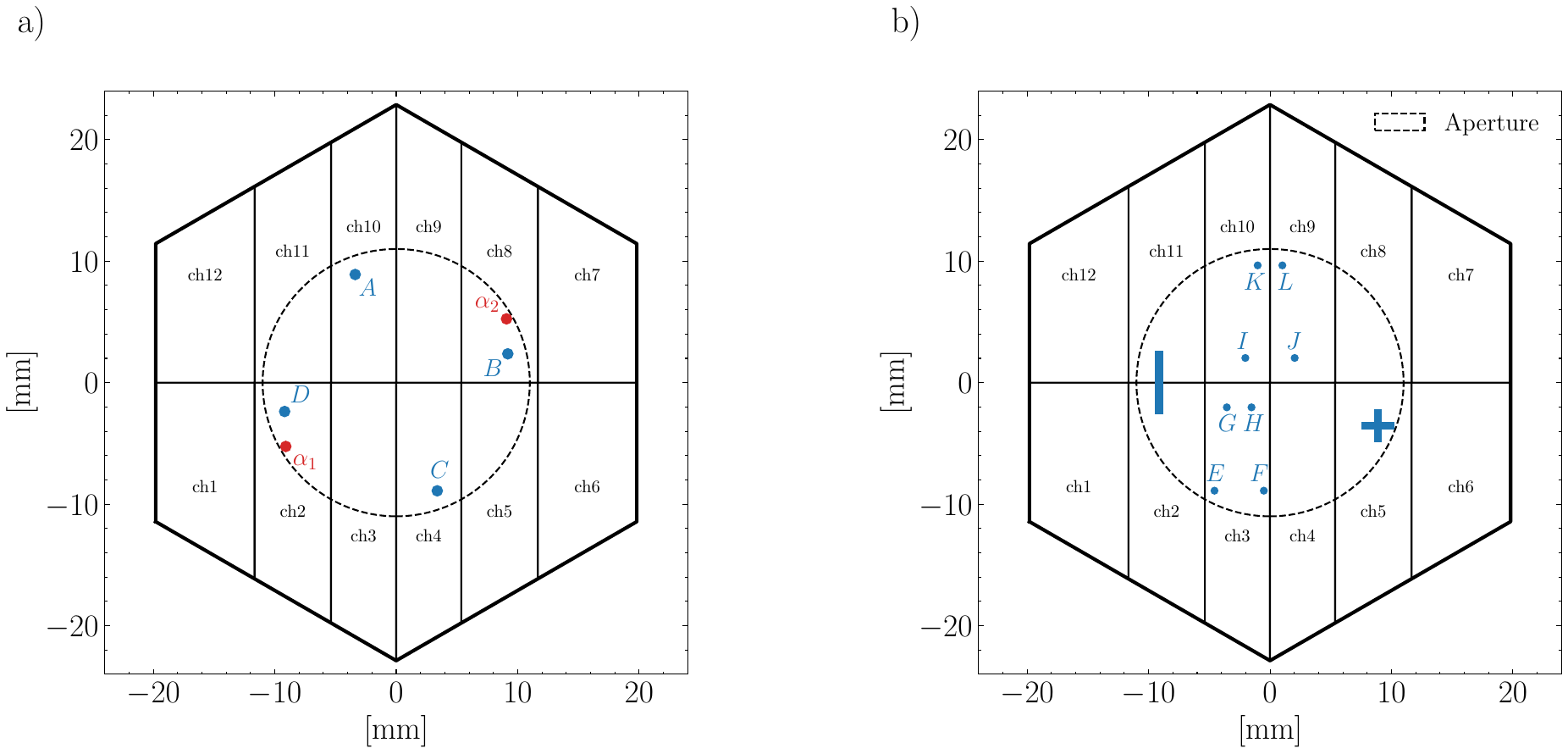}
\end{tabular}
\end{center}
\caption{Diagrams showing the layout of the anti-co prototype detector channels and the locations of openings for three different masks used in testing. (a) Geometry of the four-hole (blue A, B, C, D) X-ray mask \#1 and two-hole alpha-particle mask (red $\alpha_1$, $\alpha_2$). (b) Geometry of the 10-opening X-ray mask \#2. The dashed circle shows the extent of the aperture that was available during testing.} 
\label{fig:masks} 
\end{figure}

Two different X-rays masks were used, each consisting of \qty{0.5}{\milli\meter}-thick Cu with holes at various positions. This thickness of Cu is enough to fully block X-rays below $\sim$~\qty{30}{\kilo\electronvolt}, but provides very little attenuation at \qty{60}{\kilo\electronvolt}. We also used a similar mask behind the detector for the Am-241 alpha particle source (fully blocking alpha particles except at specified locations). Mask hole locations are shown and labeled in Figure \ref{fig:masks}. One limiting factor of 
our test setup, which was designed for Athena array testing, was its small aperture ($r \approx$ \qty{11}{\milli\meter}). We plan to deploy a larger aperture in the future, to more fully characterize the spatial response of the anti-co detector and verify its performance over the full detector area.

\section{Measurement Results}
\label{sect:results}

\subsection{Transition Properties}
\label{sect:IV}

\begin{figure}
\begin{center}
\begin{tabular}{c}
\includegraphics[width=\textwidth]{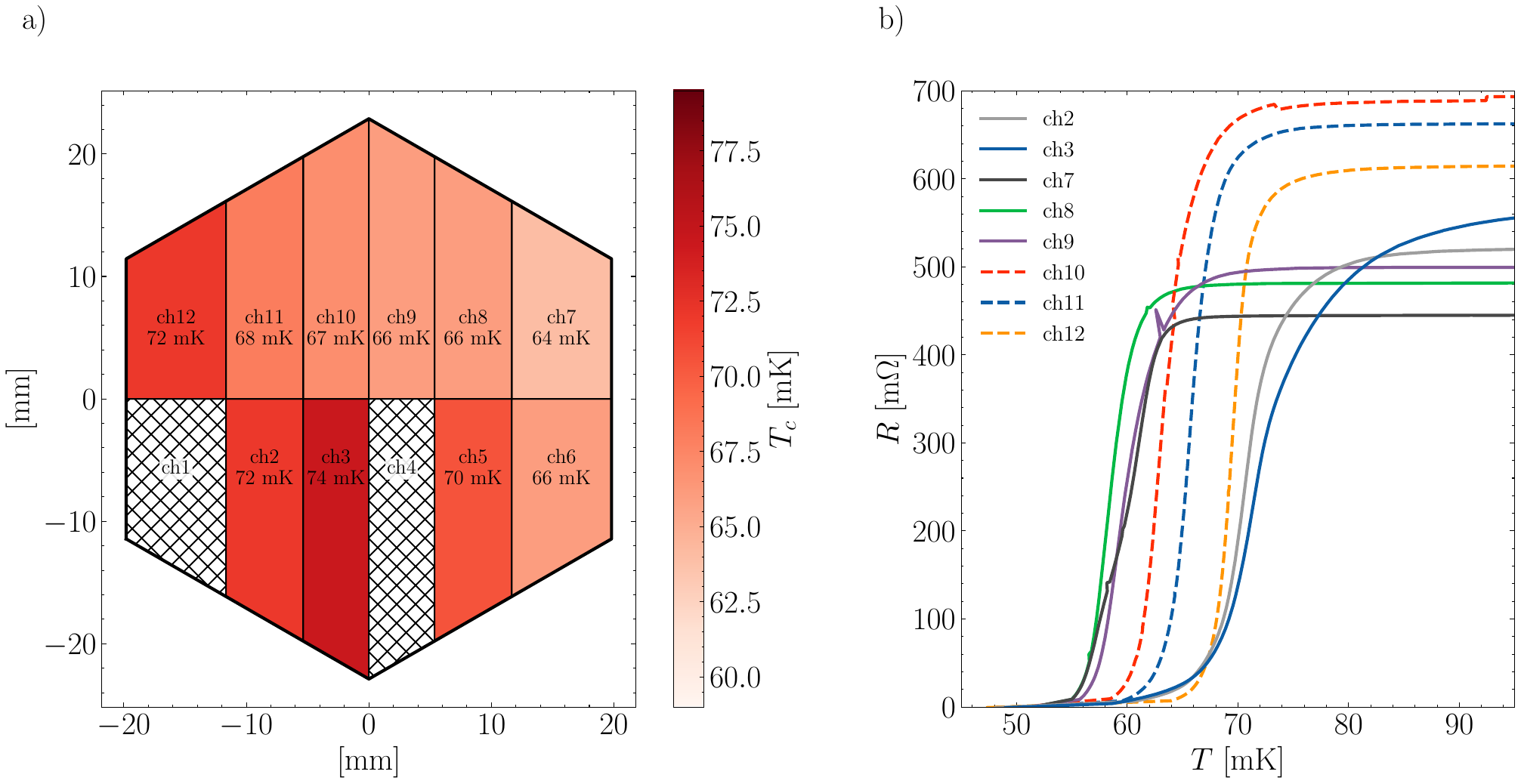}
\end{tabular}
\end{center}
\caption{(a) Heatmaps of the low TES-bias current transition temperature for the anti-co prototype. Hatched areas show inoperable channels. (b) TES resistance as a function of TES temperature for 8 of the 10 active channels. Noisy I-V data for channels 5 and 6 could not be meaningfully translated into a clean R vs. T curve. Grouping of solid and dashed lines show channels that were common-biased. A high laboratory background rate of $\sim$~\SI{0.5}{\counts\per\second} led to the various outlier points seen in the curves.} 
\label{fig:Tc_IV} 
\end{figure}

The critical temperature ($T_{\text{c}}$) for each anti-co channel was measured using a small TES-bias current, with the results shown as a heatmap in Figure \ref{fig:Tc_IV} a). As each channel is composed of a large parallel network of $\sim$ 530 TESs, the data correspond to an approximation of $T_{\text{c}}$ over the channel network. Transition properties of each channel were also evaluated by measuring the DC current response to an input TES bias voltage ($V_{\text{b}}$). The thermal conductance between each channel and the thermal bath, $G_\text{bath}$, was calculated by fitting the TES power as a function of $T_{\text{b}}$ at the midpoint of the transition ($R/R_{\text{n}}$ = 50\%). The result was $G_\text{bath}(0.1 K)$ $\sim$~\qty{200}{\pico\watt\per\kelvin}. Using this calculated value for $G_\text{bath}$, the TES resistance as a function of TES temperature was determined. The results are shown in Figure \ref{fig:Tc_IV} b). The normal resistance varies significantly between channels. Variability in the true $R_{\text{sh}}$ values is one possible source of error; however, the result is also likely related to variation in number of connected QETs in each channel due to poor-quality Al traces in the detector. 

The need to simultaneously collect data from multiple channels with different $T_c$, using a common-bias TES readout setup, meant that channels were operated at a range of operating points. We searched for bias values which allowed for simultaneous operation of all channels while still achieving reasonable gain. In practice, this resulted in operating points with $R/R_n =$~\qtyrange[range-phrase=~--~]{10}{60}{\%}. The measured $T_{\text{c}}$ spread of $\sim$15 mK across the wafer is larger than expected, given the fabrication experience of SLAC. We expect future wafers to have a $T_{\text{c}}$ spread closer to $\sim$~\qty{5}{\milli\kelvin}.

\subsection{Pulse Shapes}
\label{sect:pulses}

Energy depositions into the Si wafer produce athermal phonons that spread throughout the detector. There is an initial burst of high-energy diffusive phonons that is short lived due to rapid scattering and down conversion, giving way to a longer-lived population of lower-energy ballistic phonons \cite{Young1990}. The phonons are absorbed in QET cells and converted (via quasiparticles) to thermal energy that is measured as a current pulse in each TES. Assuming strong electrothermal feedback and negligible parasitic TES circuit resistance, the phonon energy absorbed by a TES ($E_{\text{abs}}$) is approximately equal to the energy removed by electrothermal feedback ($E_{\text{ETF}}$) and given by the following integral of the TES current pulse:

\begin{equation}
\label{eq:eabs}
E_{\text{abs}} \cong E_{\text{ETF}} = \left( 1 - 2 \frac{R_{\text{sh}}}{R_{\text{sh}} + R_0} \right) I_b R_{\text{sh}} \int \delta I_s(t) dt + R_{\text{sh}} \int \delta I^2_{s}(t)dt	
\end{equation}

\noindent where $R_{\text{sh}}$ is the shunt resistance, $R_0$ is the TES operating resistance, $I_b$ is the TES bias current, and $\delta I_s(t)$ is the change in TES current relative to the baseline value during a current pulse\cite{Irwin2005,Ren2021}. Given a parallel QET network, the measured $E_{\text{abs}}$ will be the sum of all energy absorbed by TESs within the channel; for a spatially-extended, multi-channel detector like the anti-co, one expects that depositing energy ($E_{\text{dep}}$) at one location leads to a spread of $E_{\text{abs}}$ values across all channels, trending towards lower $E_{\text{abs}}$ at further distances from the deposition location.

Considering their obvious dependence on $E_{\text{abs}}$, pulses measured in a given channel will vary based on both the magnitude and incident location of energy depositions. In general, one expects fast-rising pulses from the athermal signal, followed by a characteristic decay associated with various time constants (diffusive vs. ballistic phonon transport, TES physics, thermal links, etc). To first order, one also expects faster arrival times and larger pulse amplitudes for channels closest to the initial event location. However, the TES response is complicated by other factors. For example, large signal sizes complicate the utility of the simple model for strong electrothermal feedback in Equation \ref{eq:eabs}, increasing the nonlinearity of the response and causing $E_{\text{abs}}$ to diverge further from $E_{\text{ETF}}$. The anti-co TESs will often operate in this highly nonlinear regime, where one therefore expects pulse integrals like Equation \ref{eq:eabs} to underestimate $E_{\text{abs}}$. 

Closely related to this are the effects of local TES saturation in the area around an energy deposition. At large enough signal size, the absorbed energy is enough to drive the TES into the normal state, where it fully loses sensitivity --- this point marks the saturation energy. For a transition width $\Delta T_c$ and heat capacity $C$, the saturation energy of a TES can be estimated as $E_{\text{sat}} \approx C\Delta T_c$. With $C$ per TES of only $\sim$~\qty{0.07}{\femto\joule\per\kelvin} and $\Delta T_c$ of a few \unit{\milli\kelvin}, $E_{\text{sat}}$ for a single QET cell is on the order of \qty{1}{\electronvolt}. Over the working energy range, it is expected that the phonon energy ultimately absorbed by TESs will exceed this saturation threshold for some number of QET cells (see Section \ref{sect:modeling}), ranging from a few cells within the primary channel of incidence (local saturation) to all cells in the whole detector (full saturation) for very large energy depositions.

\begin{figure}
\begin{center}
\begin{tabular}{c}
\includegraphics[width=\textwidth]{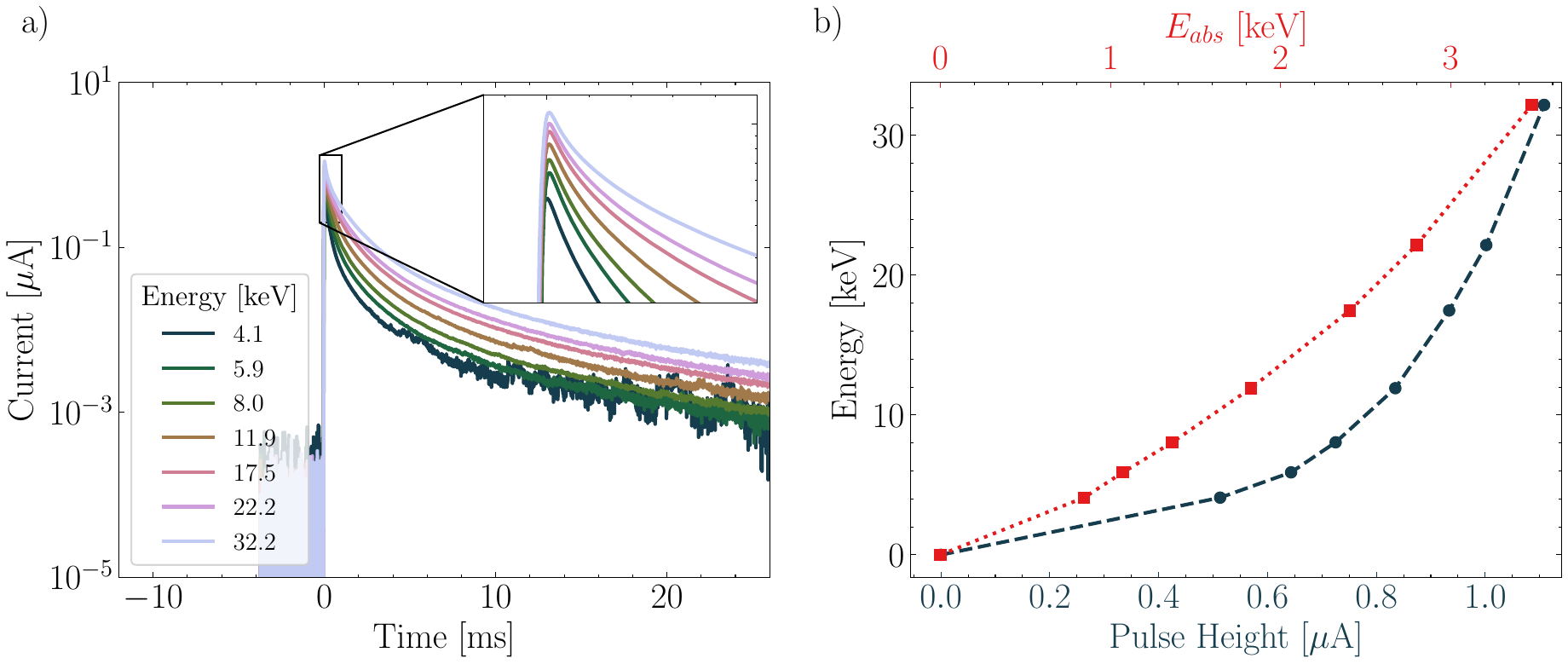}
\end{tabular}
\end{center}
\caption{(a) Average pulse shapes measured in ch10 for source position A as a function of incident X-ray energy. The inset shows the dependence of the initial fall-time with increasing energy, indicative of local saturation effects. (b) Gain plot for pulse height amplitude and $E_{\text{abs}}$ (pulse integral) for a single channel (ch10), showing the non-linearity in both. Dashed/dotted connecting lines are not fits and only visual aids.}
\label{fig:pulses} 
\end{figure}

Given that we measure each channel as a parallel network of cells, the effects of local saturation will primarily manifest as lengthened pulse decay times and non-linear energy scaling of pulse height amplitudes and pulse integrals\cite{Fink2021}. We observe these effects when exposing our anti-co prototype to X-rays at specific locations, as shown in Figure \ref{fig:pulses}. The left plot shows average pulse shapes measured in one channel (ch10) at a single mask location within that channel (region A) for a range of incident X-ray energies. As expected, the pulse height amplitude increases with event energy. However, as shown more clearly in the figure inset, the primary decay time-constant also increases with energy: from $\sim$~\qty{200}{\micro\second} at \qty{4.1}{\kilo\electronvolt} to $\sim$~\qty{300}{\micro\second} at \qty{32.2}{\kilo\electronvolt}. This behavior can be explained by local saturation of TESs; for high-energy events, the expanding $\sim$ sphere of phonons propagating from the event location in the substrate can contain sufficient energy density at the detector surface to drive a large area of QETs fully out of the superconudcing state \cite{Young1989}. The same effect can be seen in the non-linear energy scaling of pulse height amplitude (PHA) and pulse integral ($E_{\text{abs}}$) (where it is especially pronounced), as shown in the right panel of \ref{fig:pulses}. The results clearly indicate that in Equation \ref{eq:eabs}, $E_{\text{ETF}}$ is increasingly diverging from the true $E_{\text{abs}}$ as the degree of local saturation increases.

\begin{figure}
\begin{center}
\begin{tabular}{c}
\includegraphics[width=.5\textwidth]{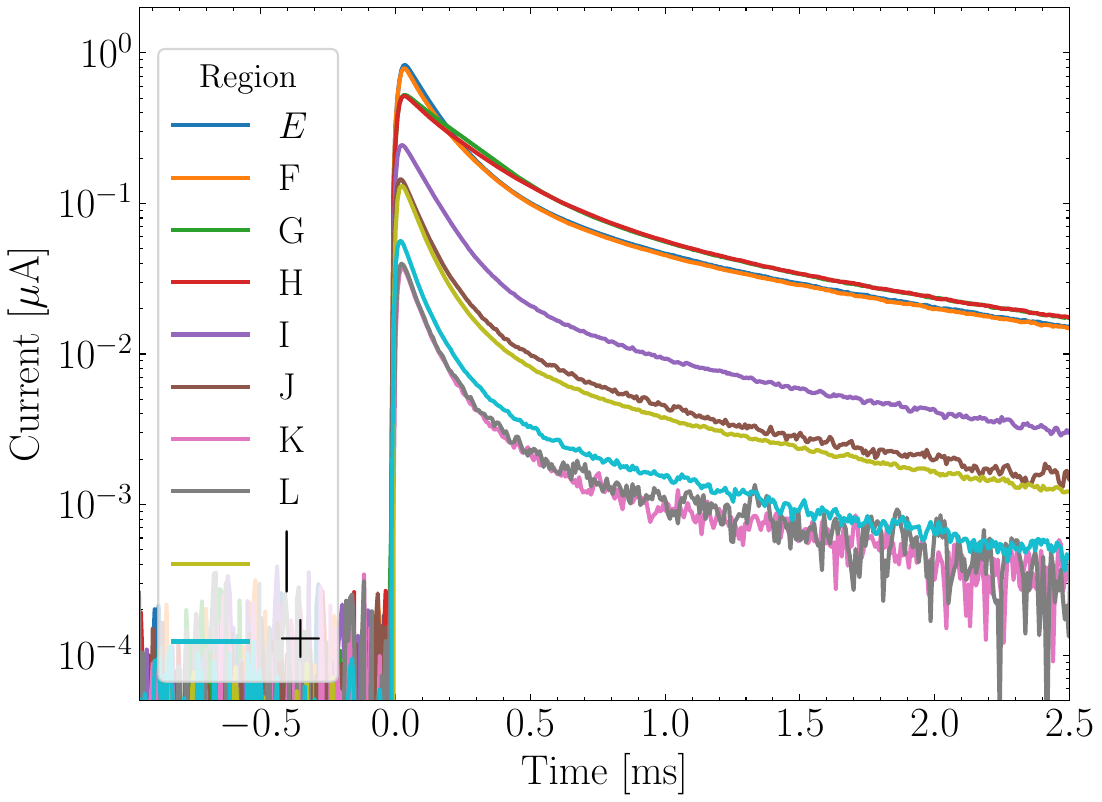}
\end{tabular}
\end{center}
\caption{Average \qty{5.9}{\kilo\electronvolt} pulse shapes measured in ch3 for all regions from X-ray mask \#2. The noticeably longer 1st falltime for events from regions G \& H compared to E \& F and is due to local saturation.}
\label{fig:pulses_region} 
\end{figure}

We also see local saturation play a role in the variation in pulse shapes between different relative positions within a channel. Figure \ref{fig:pulses_region} shows a sequence of average pulses from different mask regions measured in one channel (ch3) at a single energy (\qty{5.9}{\kilo\electronvolt}). Comparing to the map of region locations in Figure \ref{fig:masks}, we see the expected trend of longer primary fall time constant for regions that are closer to ch3, and this longer time constant generally correlates with the pulse height amplitude. An exception to this is found for regions G and H, where, compared to regions E and F, we see a lower pulse height but a longer primary fall time constant. All four of these regions are incident on ch3, however regions G and H fall further from the channel boundary than regions E and F. For a given saturation radius, a greater percentage of the saturated QET cells will therefore fall in ch3 and likewise ch3 will have a greater percentage of its QET cells saturated --- giving rise to the longer primary fall time constant. Meanwhile, the pulse height amplitude is lower for regions G and H because their positions of incidence are relatively far from the center of ch3, and therefore the total signal is spread more evenly between other neighboring channels.

\begin{figure}
\begin{center}
\begin{tabular}{c}
\includegraphics[width=.5\textwidth]{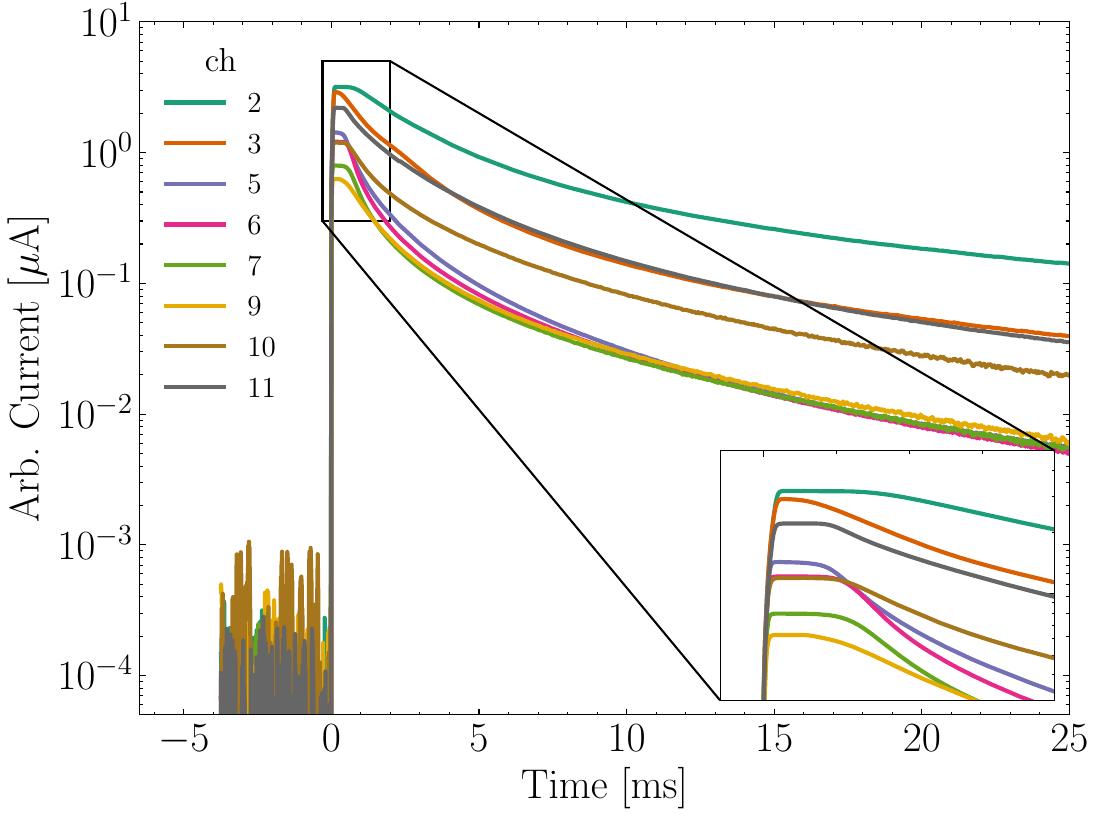}
\end{tabular}
\end{center}
\caption{Average \qty{5.5}{\mega\electronvolt} pulse shapes measured from alpha particles incident on region $\alpha_1$, demonstrating the hard saturation observed in all channels. Note that the y-axis here is arbitrary and should not be directly compared between channels.} 
\label{fig:pulses_alpha} 
\end{figure}

At large enough deposition energy, the saturation area grows to encompass the entire incident channel, and eventually even the full detector. A fully saturated channel will manifest with a flat-top pulse ceiling that we define as hard saturation. The duration of hard saturation depends on the magnitude of absorbed energy and therefore pulse integration can still be used as an energy estimator in this case, albeit not accurate to the true $E_{\text{abs}}$. Figure \ref{fig:pulses_alpha} shows average pulses measured for \qty{5.5}{\mega\electronvolt} alpha particle incident on channel 2 (region $\alpha_1$). Here every channel is hard saturated for a length of time ranging from $\sim$~\qtyrange{0.5}{1}{\milli\second}. While this hard saturation does limit the anti-co sensitivity, even at this extreme energy the dead time incurred is still acceptable for a LEM-like mission and discrimination between different types of saturated events is possible (see Sections \ref{sect:efficiency} and \ref{sect:spatial}).

\subsection{Energy Efficiency}
\label{sect:energy_eff}

In the previous section we defined the energy absorbed in a channel as a pulse integral given by Equation \ref{eq:eabs}. This absorbed energy can be compared to the known $E_{\text{dep}}$ value to define the energy efficiency:

\begin{equation}
\epsilon = \frac{E_{\text{abs}}}{E_{\text{dep}}}.
\end{equation}

Because we measure $E_{\text{abs}}$ on a per-channel basis, $\epsilon$ values are channel dependent as well. If $E_{\text{abs}}$ is summed over all channels one can define the total detector energy efficiency: 

\begin{equation}
\epsilon_{\text{det}} = \frac{\sum_{\text{ch}} E_{\text{abs}}}{E_{\text{dep}}}.
\end{equation}

\begin{figure}
\begin{center}
\begin{tabular}{c}
\includegraphics[width=.8\textwidth]{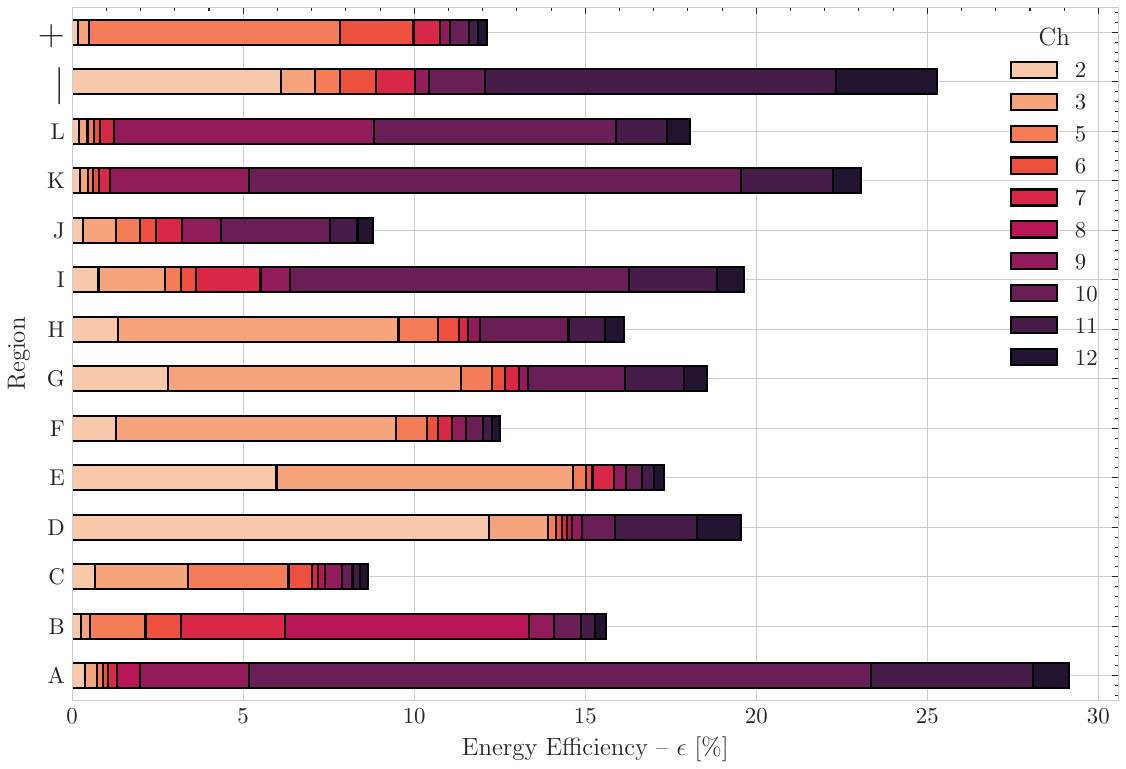}
\end{tabular}
\end{center}
\caption{Energy efficiency of the anti-co measured at 5.9 keV and for each region of the X-ray masks. For each region the full bar size shows the total detector efficiency, which is then broken down into the individual channel energy efficiency. Note that due to pulse non-linearity, inoperative channels, and some fraction of open-circuit QETs, the true energy efficiency is likely to be underestimated for many (or all) regions and this is therefore a conservative measurement.}
\label{fig:energy_eff} 
\end{figure}

Figure \ref{fig:energy_eff} shows the results of this efficiency calculation for \qty{5.9}{\kilo\electronvolt} X-rays at all mask locations. It is important to note that, as mentioned in the above section, Equation \ref{eq:eabs} tends to underestimate the true value of $E_{\text{abs}}$ --- particularly in the highly non-linear regime that some channels will reach during the measurement. Due to common channel biasing and $T_{\text{c}}$ variability, the inaccuracy is also channel dependent, contributing to the spread of values observed in the plot. The measurement is further effected by QETs that are disconnected and therefore not contributing energy to $E_{\text{abs}}$. This includes inoperable channels that we do not measure any signal from (e.g. ch4 and the corresponding $\epsilon$ deficiency in region C) as well as sub-regions of disconnected QETs within an otherwise working channel. We believe the area around region J in ch9 is likely to be an example of such a location. Due to these factors this measurement should be treated as a lower-limit on the detector energy efficiency. The highest measured detector energy efficiency of 29\% is consistent with previous measurements of efficiency for an HVeV detector of similar design \cite{Ren2021}.

\subsection{Spectral Response}
\label{sect:dE}

LEM had no requirement on the energy resolution of its anti-co, as determining the energy of incident particles is not critical for vetoing coincident events. Nevertheless, understanding the detector spectral response can be useful in obtaining a complete picture of the capabilities of the device. Given the spatially-extended nature of the detector response, the spectral and spatial component of this response are intimately intertwined, meaning that spectral response knowledge may be useful for optimization of the spatial resolution. In this section we briefly comment on the spectral performance of the anti-co prototype, while Section \ref{sect:spatial} delves into the spatial response. In the future it will likely be beneficial to more directly consider these aspects together when building an algorithm to reconstruct incident events.

\begin{figure}
\begin{center}
\begin{tabular}{c}
\includegraphics[width=.8\textwidth]{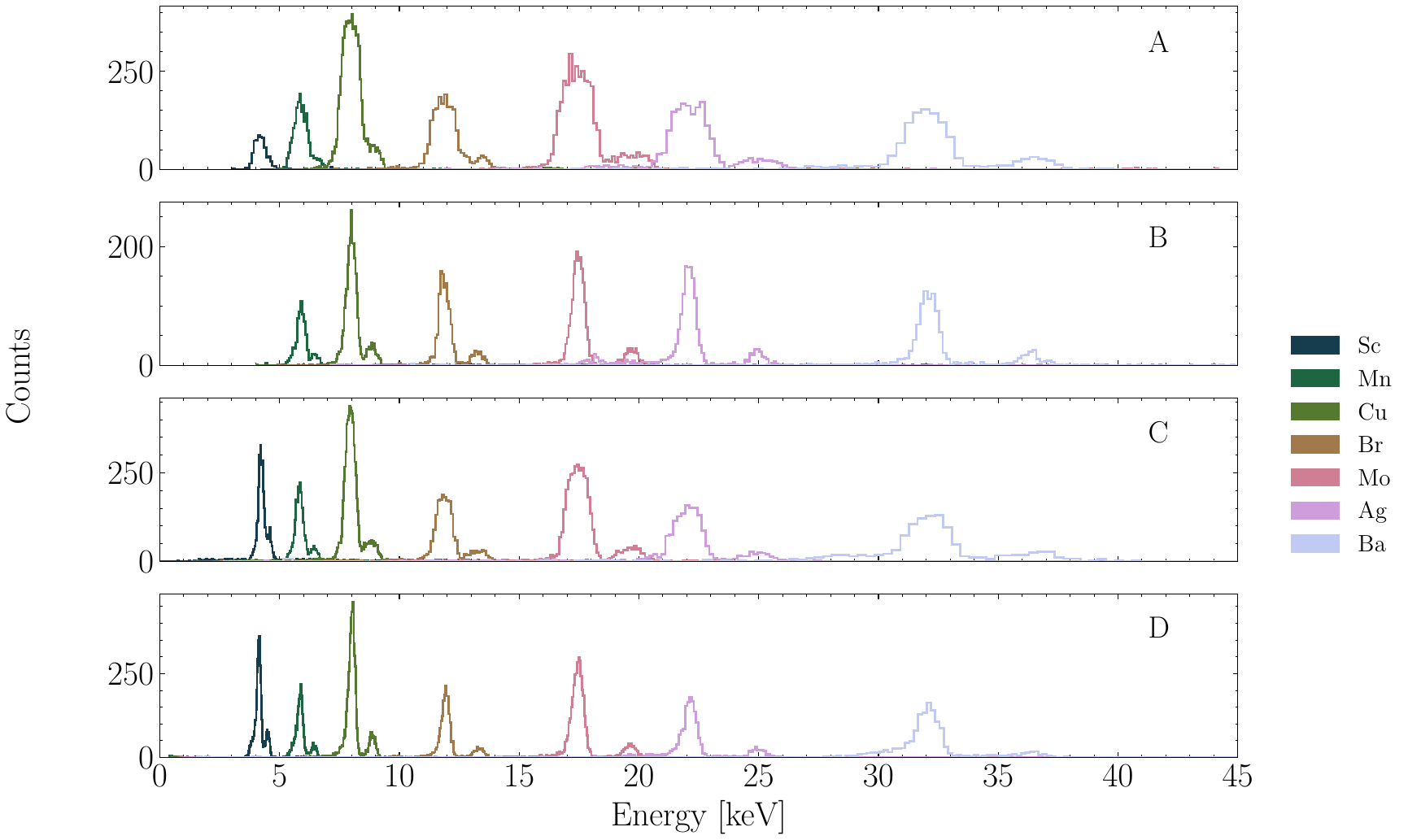}
\end{tabular}
\end{center}
\caption{Combined all-channel spectra from a number of X-ray fluorescence sources ranging from 4.1~keV to 32.2~keV, shown for each of the four locations on X-ray mask \#1 (as denoted in top-right of each subplot). These spectra show summed raw PHA histogram from all operable channels without any weighting to account for gain differences between channels. Sc is not seen in region B due to a higher triggering threshold (in nearby channel) being used than needed during this data acquisition.}
\label{fig:spectra} 
\end{figure}

If one does not spatially constrain incident energy depositions, the single-channel spectral response to a mono-energetic line source is greatly smeared in energy space by the variance in spatial response, resulting in a continuum-like response. In principle one can combine signals from all channels to ``undo'' this smearing and reconstruct the total energy for any location, but this requires high detector uniformity and/or cross-channel calibration --- neither of which are realized for the current prototype. Instead, we use the response at mask locations to investigate the spectral resolution in a more controlled manner. Figure \ref{fig:spectra} shows measured X-ray spectra for the four locations from X-ray mask \#1, using seven different x-ray fluorescence sources. These spectra show the total signal from all (operable) channels; here the total signal is calculated as a simple sum of the raw PHA from all channels, with no additional filtering or weighting applied. The centroid position of each K$\alpha$ line was found using a Gaussian fit, and then the energy gain scale for each location was evaluated by fitting a 4th order polynomial to the raw PHA centroids as a function of known X-ray energy. While the realized spectral performance is variable (primarily due to differences in channel TES operating points), the energy resolution is sufficient for rough feature separation --- including distinguishing separate K$\alpha$ and K$\beta$ lines in most cases. At \qty{5.9}{\kilo\electronvolt}, the single-location energy resolution ranges from $\sim$~\qty{200}{\electronvolt} to $\sim$~\qty{500}{\electronvolt}.

The raw PHA used above is typically not the best estimator of energy. For a standard TES microcaloriemter, energy resolution is minimized using the optimally filtered pulse height\cite{Fowler2016}. However, non-linear pulse shapes diminish the effectiveness of the traditional (Wiener) optimal filter (OF)\cite{Fixsen2014, Shank2014}. Following Ren et al., we investigated use of what they called the ``matched filter integral" (MF-Integral) energy estimator to improve the resolution in this regime\cite{Ren2021}. This approach combines pulse integration with fitting to a template. Above a selected threshold, the first part of the pulse record is directly integrated --- corresponding to the portion of the pulse with the largest degree of non-linearity. Below the threshold, where the pulse shape scales more linearly with energy, the pulse record is fit to the template and then integrated. This method can improve on simple integration by using the fit to handle the pulse tail where signal-to-noise is low. We investigated use of the MF-Integral with a varying threshold of 30\% of the average pulse height amplitude for a given energy/channel combination.

When applied to data taken with our prototype, we found that the MF-Integral improved the energy resolution in certain cases, typically corresponding to single-channel resolution for events with the largest signal-to-noise --- such as incident channels to energy depositions. However, in other cases (such as for channels far from $E_{\text{dep}}$ location) the standard OF was more successful. Because of the large range of signal sizes between channels for a single event, it is likely that some type of hybrid approach is truly optimal, but developing such an analysis chain is outside the current scope of anti-co development. 

\subsection{Threshold}
\label{sect:threshold}

As illustrated in Section \ref{sect:pulses}, the detector response is a function of event location. We define the low-energy threshold $E_{\text{th}}$ as the lowest energy at which we can detect an event in at least one channel of the anti-co. The threshold depends on event location, pulse triggering method, and sampling rate. We used a simple nearest-neighbor, finite-difference approximation as an offline trigger to evaluate the threshold for this work.  In the future, we may investigate more advanced derivative triggering algorithms involving different window sizes, such as done for X-IFU\cite{Chiarello2022}. We note that LEM was planned to operate at a higher sampling rate than was used here, which should lead to a lower anti-co $E_{\text{th}}$.

\begin{figure}
\begin{center}
\begin{tabular}{c}
\includegraphics[width=.9\textwidth]{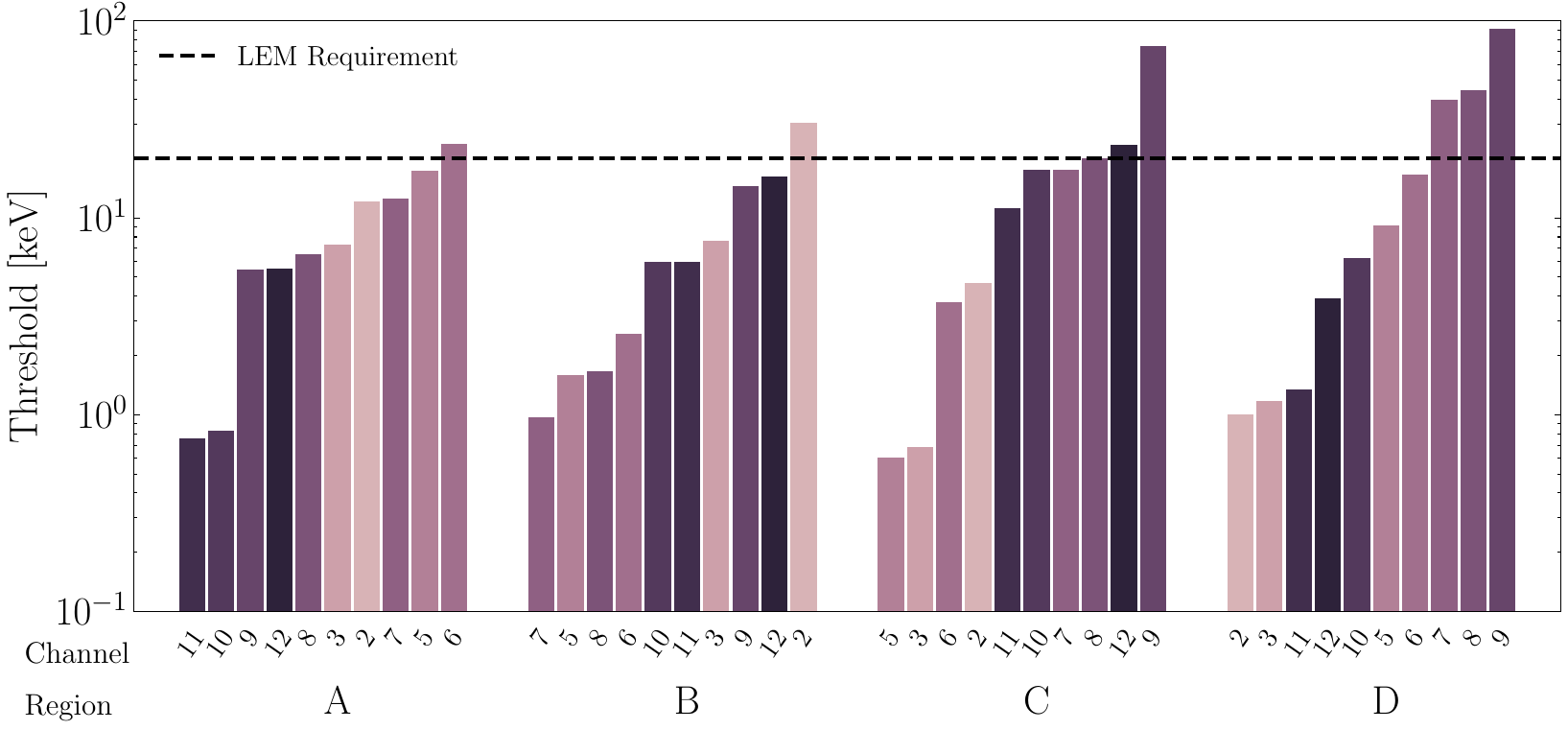}
\end{tabular}
\end{center}
\caption{The low-energy threshold of the anti-co for the ten live channels, measured as the 5$\sigma$ noise derivative of 4.1 keV X-ray events for each source location (A, B, C, D) of X-ray mask \#1. Because only a single channel anti-co trigger is needed to register an event, the anti-co detector threshold is effectively the lowest threshold of any one channel. The LEM anti-co requirement of $E_{\text{th}} < $ \qty{20}{\kilo\electronvolt}, shown here as the dashed line, is satisfied by over an order of magnitude for each location. Note that relative threshold differences between channels are partially related to different relative operating points, and therefore the shown data are not necessarily the optimally obtainable thresholds for this prototype.}
\label{fig:threshold} 
\end{figure}

The low-energy threshold was evaluated at each point for X-ray mask \#1 and the results are shown in Figure \ref{fig:threshold}. $E_{\text{th}}$ was calculated as the 5$\sigma$ value of the measured noise derivative and linearly scaled to energy-space using the lowest energy X-rays measured: 4.1~keV. Note that because of the position-dependence of the signal, different channels can be operating in different regimes (non-linear vs linear) for the same incident event energy. In effect, this means there may be an overestimation of $E_{\text{th}}$ due to the non-linearity of 4.1 keV pulses for some channels. Nevertheless, we measure a threshold below \qty{1}{\kilo\electronvolt} for all locations --- far below the \qty{20}{\kilo\electronvolt} LEM requirement.

\subsection{Detection Efficiency and Dead Time}
\label{sect:efficiency}

The detection efficiency of the anti-co is defined as the fraction of background events (with $E_{\text{dep}} > E_{\text{th}}$) incident on the detector that are triggered in at least one channel. Considering only a single incident event in isolation, this efficiency depends on pulse size and shape, system noise, and trigger method. The main limiting factor for anti-co efficiency will be the dead time induced by pile-up events. Given an average background rate $R$ and channel dead time ($t_\text{dead}$), Poisson statistics give the probability $P$ that zero events occur during the dead time window:

\begin{equation}
P = e^{-R \cdot t_\text{dead}}.
\end{equation} 

\noindent This expression also gives the fraction of detected background events. The LEM requirement of $P \ge 0.95$ can be satisfied for $t_{\text{dead}} \lesssim~$\qty{1}{\milli\second}, assuming a background rate of $R = 50$ counts/s.

We investigated the achievable dead time on the prototype with a triggering simulation that used laboratory measurements as inputs. For a range of different event energies and time intervals between pulses, the simulation generated both a primary and a secondary pulse and evaluated whether the 2nd pulse was successfully detected with our simple finite-difference derivative trigger. Pulse shape inputs were derived from measured average pulse shapes at energy calibration points; for a given simulation energy, we used a weighted average of the measured pulse shapes for the two nearest energies, with the weights set by linear interpolation of the pulse heights. We limited the maximum pulse height to the observed hard saturation limit for a given channel to account for saturation dead time for very large energies. The simulation baseline noise was also derived from laboratory data. For each simulation point we simulated 1,000 pairs of primary/secondary pulses. We then measured the fraction of primary/secondary pairs which successfully triggered the secondary pulse. 

\begin{figure}
\begin{center}
\begin{tabular}{c}
\includegraphics[width=\textwidth]{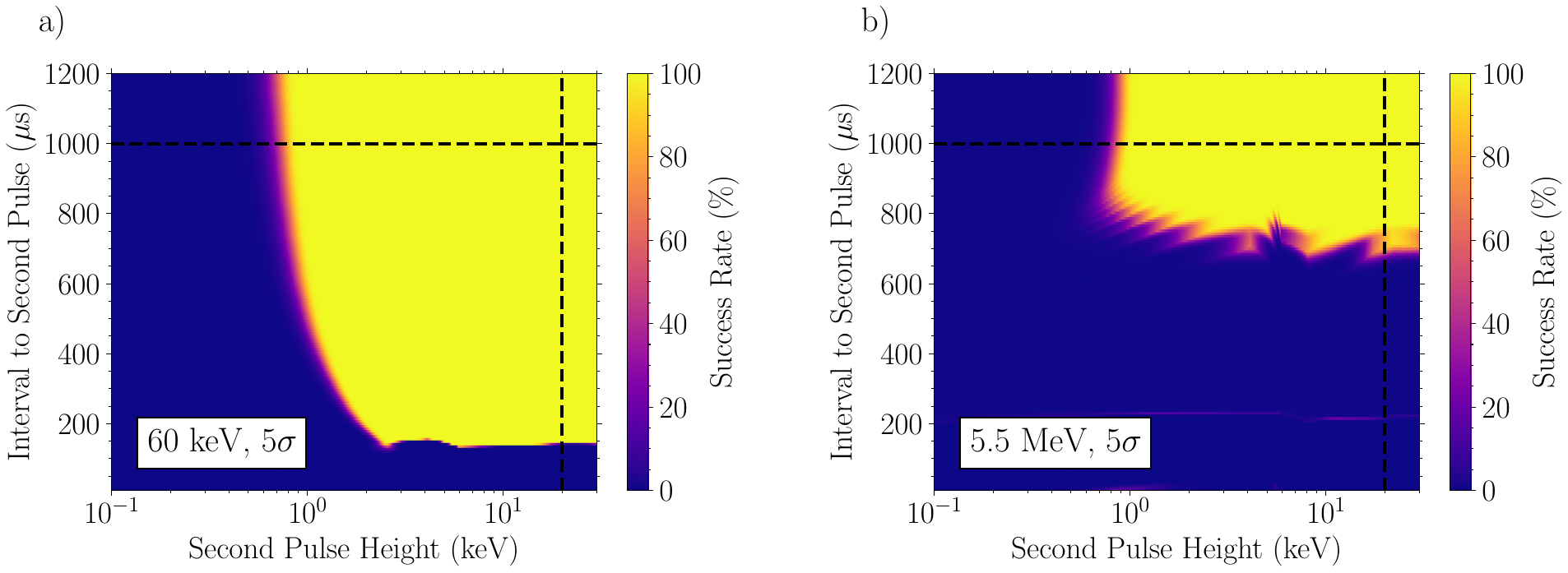}
\end{tabular}
\end{center}
\caption{Dead time simulation results for two different primary event energies: \qty{60}{\kilo\electronvolt} (a) and \qty{5.5}{\mega\electronvolt} (b). For each primary energy, a range of secondary pulse energies and time intervals are simulated, with the secondary pulse trigger success rate calculated at each point and shown as a heatmap. The dead time for a given secondary pulse energy can be considered the interval where the success rate falls below 100\%. The dashed lines show the notional LEM requirements for low-energy threshold (\qty{20}{\kilo\electronvolt}) and dead time ($\sim$~\qty{1}{\milli\second}).}
\label{fig:deadtime} 
\end{figure}

The results for two different primary event energies from $^{241}$Am are shown in Figure \ref{fig:deadtime}. These simulations were for a single channel (ch2), and used only one source location each (regions D and $\alpha_1$). Figure \ref{fig:deadtime}(a) shows that for the \qty{60}{\kilo\electronvolt} primary case, the \qty{1}{\milli\second} dead time requirement is met at a second-pulse energy threshold of $\sim$~\qty{1}{\kilo\electronvolt}. The dead time at the LEM threshold value of \qty{20}{\kilo\electronvolt} for the secondary pulse is $\sim$~\qty{0.15}{\milli\second} (equivalent to 99.3\% live time). For \qty{5.5}{\mega\electronvolt} primary events (Figure \ref{fig:deadtime}(b)), we move into the hard saturation regime and therefore the dead time is significantly extended to $\sim$~\qty{0.75}{\milli\second} (96.3\% live time) --- yet even this result meets the baseline LEM requirements. The whole-detector dead time will be a complicated function of energy and position of both the primary and secondary events. While a full analysis has not been completed at this time, the results shown above --- particularly the \qty{5.5}{\mega\electronvolt} case --- approximates a worst-case scenario. 

\subsection{Spatial Resolution}
\label{sect:spatial}

Unlike traditional pixelated detectors, the spatial resolution of the anti-co is related to both the channel number and layout as well as the details of athermal phonon diffusion through the substrate. Discriminating between events that occur in different locations of the anti-co is not required to achieve LEM's primary goals. However, such a capability would allow one to reconstruct the tracks of cosmic-rays passing through the detector and therefore determine the corresponding incident position on the main detector array. When an event is detected in both the main array and anti-co, there will be a corresponding coincidence-induced deadtime: a time window during which main array events must be discarded. With the benefit of anti-co spatial resolution, this coincidence-induced deadtime can be applied only to events within a small spatial area, thereby increasing the instrument throughput. Fully characterizing the anti-co spatial resolution and/or corresponding main array veto scheme is outside the scope of this work. Nevertheless, below we summarize some of our findings to-date that could impact future optimizations along this axis.

\begin{figure}
\begin{center}
\begin{tabular}{c}
\includegraphics[width=\textwidth]{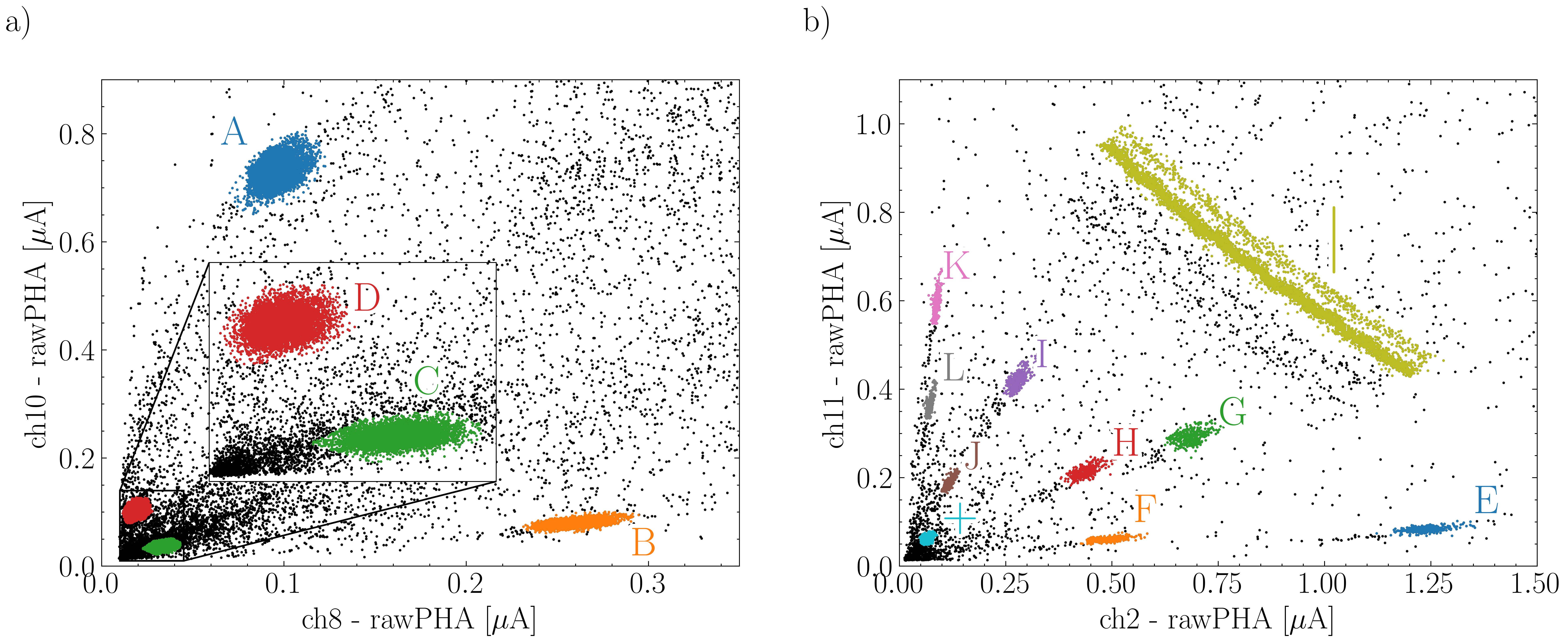}
\end{tabular}
\end{center}
\caption{Scatter plots of the pulse height amplitude (PHA) for X-ray pulses in one anti-co channel vs. another anti-co channel using (a) X-ray mask \#1 with a Cu X-ray fluorescence source, and (b) X-ray mask \#2 with a Br fluorescence source. In each case, groupings of points can be identified (and are colored) corresponding to specific regions on the masks, the identification of which was completed manually via the analysis of many such plots. The relatively high laboratory background rate of $\sim 0.5$~counts/s accounts for the large number of unlabeled scatter points on the plots. ``Shadow" event clusters on plot (b) at energies just below the Br features are due to background Cu fluorescence events.}
\label{fig:regions_pha} 
\end{figure}

In describing the differences in response between different locations (Section \ref{sect:pulses}) we have relied on the inherent ability of the anti-co to discriminate between different hole locations on the same mask. At the coarsest level, event position resolution over the channel-size scale is demonstrated from testing with X-ray mask \#1, while sub-channel resolution can be derived using X-ray mask \#2. In principle, several different pulse properties can be used to differentiate between events at different locations, including, e.g., those mentioned in Section \ref{sect:pulses} (PHA, pulse integral, $\tau$), as well as pulse arrival time, rise time, etc. However, we found that using just a single metric, such as raw PHA, is already sufficient to demonstrate relatively high resolving ability. Figure \ref{fig:regions_pha} shows scatter plots of raw PHA in one anti-co channel vs. PHA in another anti-co channel for collimated low-energy X-ray events. The dense clusters of points in each plot are labeled by their corresponding X-ray mask hole as defined in Fig. \ref{fig:masks}. These regions were manually identified by looking at a number of such PHA-PHA plots and correlating the relative position of clusters within the plots to mask regions. Note that because channel-to-channel cross-calibrations for raw PHA were not done, only relative (not absolute) comparisons of PHA values between channels are appropriate. 

Figure \ref{fig:regions_pha} demonstrates that even using only one parameter on a single pair of channels shows clear separation of all tested regions. This strongly suggests the anti-co offers spatial resolution  of $\le$~\qty{2}{\milli\meter}, the minimum spacing between mask holes. Some discrimination ability persists even for channels far from the primary event location; for example, all four mask regions (E, F, G, H) inside ch3 can be resolved using only ch2 and ch11. Moreover, the observation of distinctive clusters of events corresponding to extended mask features, e.g., the slot and cross regions, demonstrates the capability for sub-mm spatial resolution. This precision could be improved even further by simply studying multiple channels or pulse features simultaneously.

\begin{figure}
\begin{center}
\begin{tabular}{c}
\includegraphics[width=\textwidth]{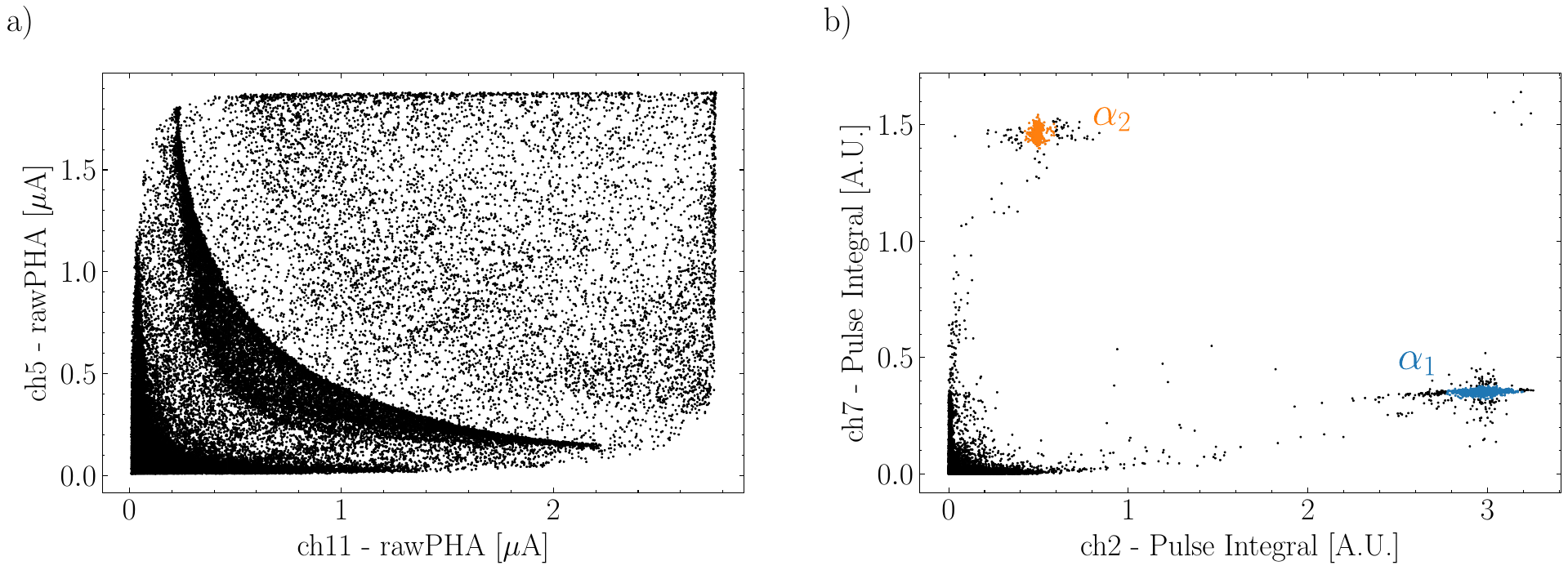}
\end{tabular}
\end{center}
\caption{(a) PHA-PHA scatter plot for two masked channels (ch5, ch11) with a 60 keV X-ray source that is not stopped in the mask. The primary ``banana" structure seen at the highest PHA values arises from 60 keV X-ray events occurring at different distances from each channel. (b) Pulse integral scatter plot for two channels (ch2, ch7) with a 5.5 MeV alpha-particle source that is masked except at two locations. Even though the pulses are hard-saturated, the event locations can be distinguished via pulse integral analysis.}
\label{fig:regions_highE} 
\end{figure}

As shown on the two-dimensional PHA spectrum of Figure \ref{fig:regions_highE}~(a), \qty{60}{\kilo\electronvolt} X-rays from an Am-241 X-ray source that pass through our mask produce a distinctive ``banana" type structure at large pulse-heights that stretches out to the two axes; each end of this feature corresponds to \qty{60}{\kilo\electronvolt} X-rays that deposit most of their energy in just one of the two channels analyzed (ch5 and ch11). The ``dip" in total PH near the center of the banana-shaped feature is caused by \qty{60}{\kilo\electronvolt} X-rays that deposit significant energy elsewhere in the detector. The similar ``banana" structure shown at lower PHA values is due to a combination of low-energy Am-241 X-ray lines and Cu florescence events.

In the regime of hard-saturation, a solely PHA-based approach losses effectiveness. However, the pulse integral method still can be used effectively to correlate event energy and position. Figure \ref{fig:regions_highE}~(b) demonstrates this approach for \qty{5.5}{\mega\electronvolt} alpha particle events; clearly, the two alpha-mask hole locations can be distinctly identified in pulse-integral space. This result is promising for retaining some degree of spatial resolution under the case of hard-saturation, in addition to underscoring the utility of using multiple pulse properties when evaluating events. Developing an algorithm to efficiently use all of this information will be important in the long term. We have started to perform algorithmic testing and are investigating various two-dimensional fits and basic machine learning techniques. While the results show promise, further development is needed to yield robust results in this area. 

\section{Modeling}
\label{sect:modeling}

The anti-co model is being developed using G4CMP\cite{Kelsey2023}, an add-on to the common Geant4 simulation package\cite{Allison_2016}. G4CMP simulates phonon and charge transport physics and has been successfully used in a variety of cryogenic sensor applications --- including SuperCDMS HVeV detectors that utilize QET cells similar in design to those comprising our anti-co detector. During initial anti-co model development work, G4CMP used a quasiparticle-trapping algorithm less advanced than the one available now. As a result, an early simulation-tuning effort for the HVeV detector response produced a well-fit detector response, but yielded non-physical results for the superconducting energy gap of aluminum. In response, we chose physically motivated values for all model parameters without attempting to tune to experimental results, since full simulation-tuning capability using raw data would require knowing, e.g., the exact QET yield per channel on the prototype device. Initial simulation results are thus not expected to exactly match the real device at this time, but are useful as a general aide in understanding detector response and for establishing a foundation from which to further develop the model. 

\begin{figure}
\begin{center}
\begin{tabular}{c}
\includegraphics[width=\textwidth]{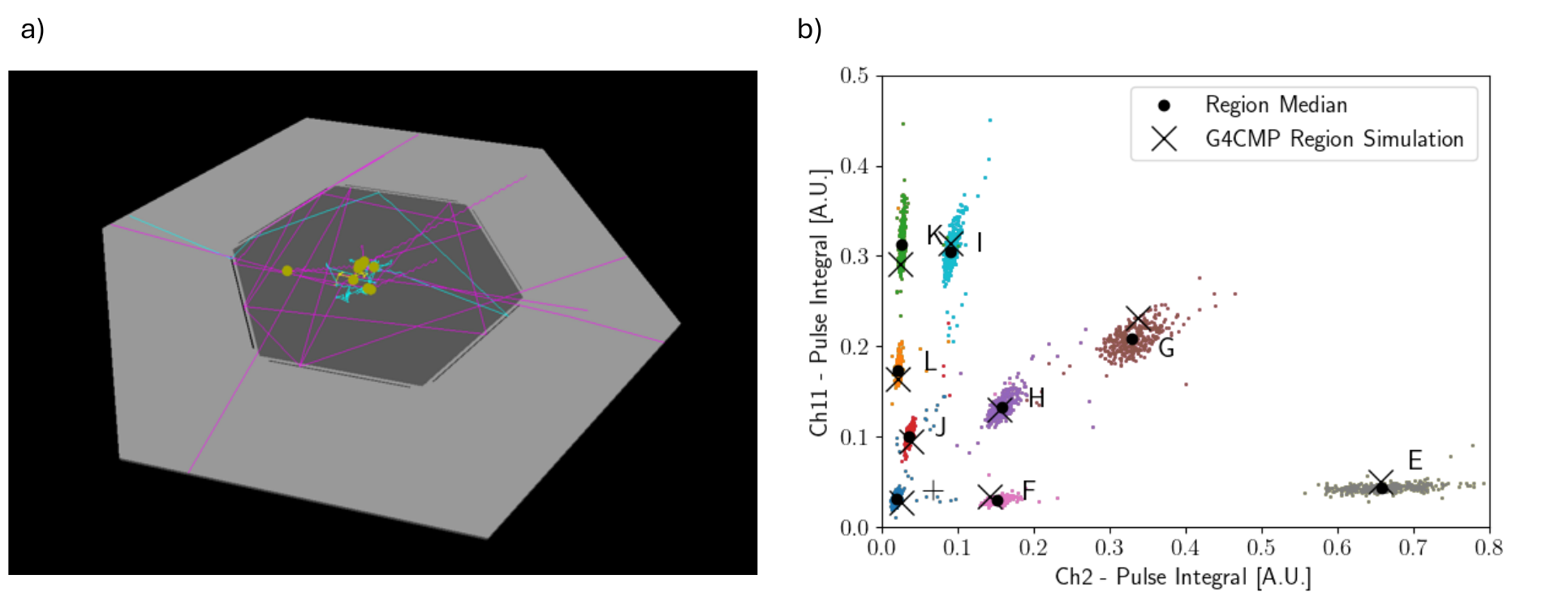}
\end{tabular}
\end{center}
\caption{(a) Geometry and example phonon tracks of the G4CMP anticoincidence Si detector model. The extent of the Si wafer is shown in light gray. The Al/W QET elements that comprise the 12 independent phonon-sensing, quasiparticle-trapping channels of the anti-co are shown in dark gray. Isolation trenches etched along the edges of the 12-channel region are included to minimize phonon loss to the wafer exterior. Cyan, magenta, and bright-yellow tracks denote fast transverse, slow transverse, and longitudinal phonon modes, respectively. Dark yellow circles mark where simulated phonons were successfully absorbed, converted into quasiparticles, and detected as an energy pulse. (b) Comparison of the G4CMP simulation results to measured data for low-energy X-rays. The plot shows a pulse-integral scatter plot between ch2 and ch11 using X-ray mask \#2. The simulation was calibrated using a large array of precisely located, G4CMP-generated \qty{6}{\kilo\electronvolt} primary events. A quadratic fit was applied to the measured vs. simulated energy absorbed for each channel in order to account for non-linear detector effects --- see text for details.}
\label{fig:model}
\end{figure}

The geometry of the $\langle$100$\rangle$ Si wafer was implemented in G4CMP and a hexagonal detector region was set to cover the area of the 12 channels. A 15\% phonon collection efficiency was set as a first order approximation to the energy-collecting Al ``fin" coverage area. Event energy was deposited at specific locations within the 500 micron-thick wafer. Tracks from the resulting fast transverse, slow transverse and longitudinal simulated phonons were mapped in 3-dimensions to determine where phonons were absorbed in the Al fins, and thus where they created quasiparticles that could diffuse into the TES of a QET and contribute energy $E_{\text{abs}}(x,y)$ to a pulse.
Figure \ref{fig:model} (a) shows a visualization of several simulated phonon-induced events for the given detector geometry. 

We explored the capability of the model to broadly reproduce observed features in the data. Calculating $E_{\text{abs}}(r)$ using the model, where $r$ is the radial distance from the initial event location to the individual QET where the energy was deposited, supports the idea that we expect $\sim$ 1 mm-scale or larger local TES saturation radii for many events. As described in Section \ref{sect:pulses}, we can roughly estimate the QET saturation energy as $\sim$~\qty{1}{\electronvolt}, or \qty{5}{\electronvolt\per{\square{\milli\meter}}} given a QET cell effective area of of \qty{0.2}{\square\milli\meter}. Simulation results predict a \qty{6}{\kilo\electronvolt} energy deposition gives rise to an $E_{\text{abs}}$ density distribution that exceeds this value out to $r \approx$~\qtyrange{1}{2}{\milli\meter}. This is consistent with observations of partial local saturation of channels, and with observations of the degree of partial-channel saturation varying when moving deposition locations closer or further from channel boundaries on the mm-scale. 

We also compared simulation results to experimental data with masked X-rays. As an example, Figure \ref{fig:model} (b) shows a pulse-integral scatter plot for the low-energy regime for ch2 vs. ch11, where each cluster of points in a given color corresponds to a different collimator hole in X-ray mask \#2. Also shown on the plot are larger points that mark the geometric median of each grouping of measured points, in addition to crosses that mark the results of G4CMP simulations for the same conditions. Note that the G4CMP simulations output results in $E_{\text{abs}}$, while the measurements are of the $E_{\text{ETF}}$ pulse integral, which will not necessarily align --- especially when affected by variable channel bias points, local saturation, and reduced QET fabrication yield. To account for these effects, a 2nd order polynomial was fit to $E_{\text{abs}}$ (simulation) vs. $E_{\text{ETF}}$ (measured) for each channel separately, thereby putting the results on the same arbitrary scale, as shown in the plot. The resulting fits demonstrate that the model can reproduce results that are broadly consistent with individual region locations within pulse-integral space, which gives us some confidence that the model is successfully capturing the physics of phonon transport and quasiparticle absorption in the detector. For this analysis, the long, rectangular hole in mask \#2 that extends over both channels in this study was ignored due to large nonlinear effects that a simple quadratic fit was not equipped to handle. 

In the future, we plan to update the model to use the latest G4CMP version and therefore take advantage of improved quasiparticle algorithms. Our plans also include updating the model geometry to resolve individual QET geometries, which may help to provide a deeper understanding of local QET saturation and therefore the spatial dependence of pulse shapes. With more testing from a higher-fabrication-yield detector, it should be possible to tune simulation parameters to better reproduce experimental data. A fully-tuned model will enable more detailed analysis, optimization, and calibration of the anti-co detector. One avenue where this may be useful is for evaluating different channel layouts; we currently believe a concentric-type channel will be optimal for position sensitivity and a variety of designs are under consideration for future devices. In addition, a better-tuned model will enable more precise position-grid data to be generated, enabling more efficient and more accurate calibration of real anti-co detectors.

\section{Conclusion}
\label{sect:conclusion}

We have presented the first characterization results for a full-scale anti-co detector designed for the LEM mission concept that includes 12 channels and more than 6300 QET cells over a \qty{14}{\square{\centi\meter}} area. Measurements of this prototype imply a low-energy threshold below \qty{1}{\kilo\electronvolt}, a live-time fraction of 96\% or better at the nominal count rate, and spectral resolution as good as $\sim$~\qty{200}{\electronvolt} at \qty{5.9}{\kilo\electronvolt}. If operated in concert with the main LEM array, the measured performance indicates the anti-co would be capable of robustly tagging coincident events and therefore significantly lowering LEM's non X-ray background to meet the requirement of $<$ \qty{2}{\counts\per{\second\per{\kilo\electronvolt\per{\FoV}}}}. We also investigated the spatial response and associated resolution of the detector, with results implying mm-scale resolution or better can be achieved once a suitable reconstruction algorithm is developed. This capability has the potential to reduce the throughput loss of the main array during coincidence-induced dead time, but more work is needed in this area; there is likewise a need to define and develop the specific method by which events will be mapped to the main array to determine coincidences. 

We also outlined a summary of the current status of modeling and simulation efforts for the anti-co detector. While currently untuned and limited in scope, the anti-co model has aided understanding of pulse features like local QET saturation and is able to broadly reproduce trends in the experimental data. Future model development and tuning will increase fidelity and enable more detailed studies. We plan to continue development efforts for our anti-co design and evolve it to higher-fidelity prototypes, including both the W-TES that was the focus of this work as well as the Mo/Au-based design. Future iterations will include using wider superconducting traces in the interior of the array in order to improve QET fabrication yield, at the expense of a possible small efficiency loss due to the additional Al. We are also considering other design modifications, such as different channel layouts to optimize spatial resolution, and modified TES geometries to improve dynamic range.

\subsection* {Acknowledgments}

The material is based upon work supported by NASA under award number 80GSFC21M0002. 

\bibliography{report}   
\bibliographystyle{spiejour}   

\listoffigures
\listoftables

\end{document}